\documentclass[sigconf,screen]{acmart}
\acmConference[ICSE 2024]{46th International Conference on Software Engineering}{April 2024}{Lisbon, Portugal}
\usepackage{textcomp}
\usepackage{xcolor}
\usepackage{url}

\usepackage{subfigure}
\usepackage{booktabs}
\usepackage{balance}

\usepackage{graphicx}
\usepackage{algorithm}
\usepackage{algorithmic}
\usepackage{xspace}
\usepackage{multirow}
\usepackage{soul}

\newcommand{\etal}{{\em et al.}\xspace}
\newcommand{\ie}{{\em i.e.},\xspace}
\newcommand{\eg}{{\em e.g.},\xspace}
\newcommand{\methodname}{DivLog\xspace}

\newcommand{\modify}[1]{\textcolor{black}{#1}}

\def\BibTeX{{\rm B\kern-.05em{\sc i\kern-.025em b}\kern-.08em
    T\kern-.1667em\lower.7ex\hbox{E}\kern-.125emX}}

\AtBeginDocument{%
  \providecommand\BibTeX{{%
    Bib\TeX}}}

\copyrightyear{2024}
\acmYear{2024}
\setcopyright{acmlicensed}\acmConference[ICSE '24]{2024 IEEE/ACM 46th
International Conference on Software Engineering}{April 14--20, 2024}{Lisbon,
Portugal}
\acmBooktitle{2024 IEEE/ACM 46th International Conference on Software
Engineering (ICSE '24), April 14--20, 2024, Lisbon, Portugal}
\acmPrice{15.00}
\acmDOI{10.1145/3597503.3639155}
\acmISBN{979-8-4007-0217-4/24/04}

\begin{document}


\title{\methodname: Log Parsing with Prompt Enhanced In-Context Learning}

\author{Junjielong Xu}
\email{junjielongxu@link.cuhk.edu.cn}
\affiliation{%
    \institution{School of Data Science, The Chinese University of Hong Kong, Shenzhen (CUHK-Shenzhen), China}
    \country{}
}
\author{Ruichun Yang}
\email{ruichunyang@link.cuhk.edu.cn}
\affiliation{%
    \institution{School of Data Science, The Chinese University of Hong Kong, Shenzhen (CUHK-Shenzhen), China}
    \country{}
}
\author{Yintong Huo}
\email{ythuo@cse.cuhk.edu.hk}
\affiliation{%
    \institution{Dept. of Computer Science \& Engineering, The Chinese University of Hong Kong, China}
    \country{}
}
\author{Chengyu Zhang}
\email{dale.chengyu.zhang@gmail.com}
\affiliation{%
    \institution{ETH Zurich,}
    \country{Switzerland}
}
\author{Pinjia He}
\authornote{Pinjia He is the corresponding author.}
\email{hepinjia@cuhk.edu.cn}
\affiliation{%
  \institution{School of Data Science, The Chinese University of Hong Kong, Shenzhen (CUHK-Shenzhen), China}
  \institution{Shenzhen Research Institute of Big Data, China}
\country{}
}

\begin{abstract}

Log parsing, which involves log template extraction from semi-structured logs to produce structured logs, is the first and the most critical step in automated log analysis.
However, current log parsers suffer from limited effectiveness for two reasons.
\textit{First}, traditional data-driven log parsers solely rely on heuristics or handcrafted features designed by domain experts, which may not consistently perform well on logs from diverse systems. 
\textit{Second}, existing \modify{supervised} log parsers require model tuning, which is often limited to \modify{fixed} training samples and causes sub-optimal \modify{performance} across the entire log source.
To address this limitation, we propose \textit{\methodname}, an effective log parsing framework based on the in-context learning (ICL) ability of large language models (LLMs).
Specifically, before log parsing, \methodname samples a small amount of offline logs as candidates by maximizing their diversity.
Then, during log parsing, \methodname selects five appropriate labeled candidates as examples for each target log and constructs them into a prompt.
By mining the semantics of examples in the prompt, \methodname generates a target log template in a training-free manner.
In addition, we design a straightforward yet effective prompt format to extract the output and enhance the quality of the generated log templates.
We conducted experiments on 16 widely-used public datasets.
The results show that \methodname achieves (1) 98.1\% Parsing Accuracy, (2) 92.1\% Precision Template Accuracy, and (3) 92.9\% Recall Template Accuracy on average, exhibiting state-of-the-art performance.

\end{abstract}



\maketitle

\section{Introduction}\label{sec:intro}

\begin{figure}[htpb]
\centering
\includegraphics[scale=0.58]{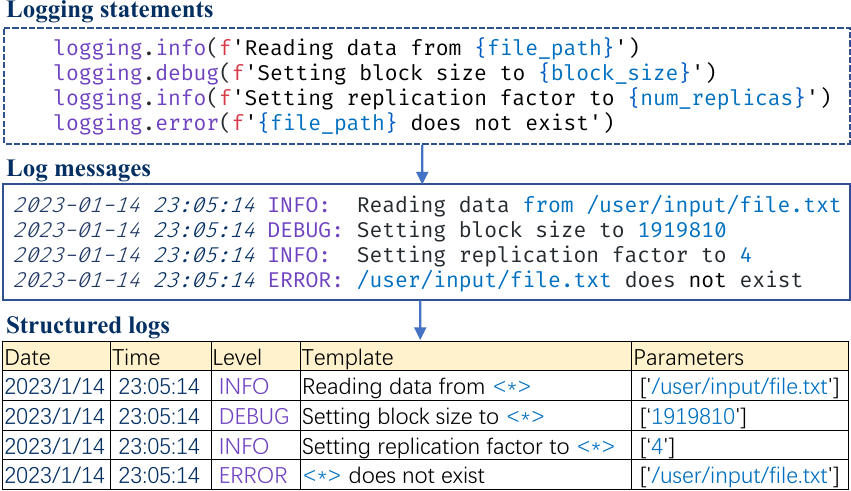}   
\caption{A simple example of log parsing.
The logging statements are typically not accessible in industrial scenarios.}
\label{fig:parsingexample}
\end{figure}

Modern software systems, including online services such as Google Search and Bing Search, and system software such as Android and Windows, have become an essential part of our lives, serving millions of users globally.
These systems produce valuable software logs continuously, providing a rich resource for maintainers to perform downstream tasks, such as anomaly detection~\cite{deeplog,exp,mininglog,robustlog}, root cause analysis~\cite{localizelog,idlogm,idlogc}, and program verification~\cite{log2,logverify}. 
The first step of log analysis is log parsing, \ie converting semi-structured log messages into structured log messages. 
Manual log parsing is impractical due to the enormous volume of logs generated~\cite{survey}.
Therefore, numerous data-driven automatic log parsers have been proposed, including traditional unsupervised parsers~\cite{ael,spell,logram,drain} and \modify{deep learning (DL) based} supervised parsers~\cite{uniparser,semparser,logppt}.
Existing log parsers primarily distinguish between \textit{constants} and \textit{variables} in a log message without the guide of logging statements.
As shown in Fig.~\ref{fig:parsingexample}, constants are the tokens written by developers in the logging statements (\eg a description of a software operation), while variables are tokens that record run-time environments (\eg a directory path).
These constants make up a \textit{log template}, while variables are treated as \textit{parameters}.

However, existing log parsers still face limited robustness~\cite{benchmark,logppt}, leading to unsatisfactory accuracy in diverse logs for two reasons:

\textit{First}, unsupervised log parsers utilize specially designed features or heuristics (\eg n-gram~\cite{logram}, prefix tree~\cite{spell,drain}) based on domain knowledge to extract common patterns for log parsing.
As a result, they often fall short in log sources whose template design does not match well with the handcrafted features.
For example, the most widely-used parser, Drain,~\cite{drain} assumes that leading tokens are constants, and logs from the same template share the same length.
However, they are inadequate for logs \modify{from \texttt{E8} event} in Proxifier~\cite{benchmarkrepo}, \eg \texttt{"ss.bdimg.com:80 close, 89652 bytes (87.5~KB) sent, 599249 bytes~(585~KB) received, lifetime 37:51"}, where the prefix is a variable \modify{(\ie URL)} and log length is flexible due to the optional tokens in brackets (\ie no bracket tokens when the transferred bytes less than 1KB).
\modify{Moreover, unsupervised parsers usually need preprocessing~\cite{survey}, wherein regex replace common variable tokens with the wildcard \texttt{"<*>"}.
Developers typically invest extra effort in designing distinct regex for different datasets with their domain knowledge due to the significant impact of regex quality on parsing performance~\cite{evaluationstudy}, which further limits their usability.
}

\textit{Second}, existing supervised log parsers typically need to train models to mine the data characteristics in target log samples~\cite{semparser,logppt}.
\modify{This training process may restrict the models to the specific training data, leading to sub-optimal generalization performance across the overall dataset, especially when the dataset exhibits significant sample diversity.}
\modify{For instance, the state-of-the-art parser, LogPPT~\cite{logppt}, extracts a few of the most frequent variable tokens from the adaptively selected logs to generate virtual labels for model training.}
\modify{However, for diverse and complex datasets such as Mac~\cite{benchmarkrepo}, there might be no representative logs or variables for overall dataset, as the dataset contains large variance of logs and variables (\eg \texttt{E253} event only contains one token without variable, while \texttt{E277} event contains 78 tokens with 36 variables) and imbalanced template distribution (\eg the number of logs in each template ranging from 1 to 166, with over 50\% of templates containing only one log).}
Overall, there is a pressing need for a more effective log parsing technique that matches the following requirements: (1) Sufficiently automated to reduce the human effort and domain knowledge involved. (2) Demonstrates decent results on complex data and ensures stability across various log datasets.


To this end, this paper proposes \methodname, an effective and training-free log parsing framework without requiring human effort in feature design or hyperparameter tuning.
\modify{\methodname adopts in-context learning (ICL) ability of large language models (LLMs) ~\cite{promptsurvey} to generate log templates via mining the common features of related log examples in the prompt.}
Specifically, \methodname employs GPT-3~\cite{gpt3} as the backbone, the first LLM that exhibits the ability to learn semantics in prompt contexts.
\modify{Before log parsing, \methodname selects a fixed amount of logs by explicitly maximizing their diversity to build a candidate set for subsequent prompting.}
\modify{During the parsing process, it retrieves the top five most relative log examples and their labels from candidate set for each target log based on their similarity.
Then, \methodname organizes them in a special prompt format, which is designed to enhance ICL performance for log parsing.
With the guidance of examples and their labels in the prompt, \methodname generates log templates without necessitating model tuning.}
\methodname has been evaluated on 16 public datasets from LogPAI.~\cite{benchmark}
The results show that \methodname achieves (1) 98.1\% Parsing Accuracy, (2) 92.1\% Precision Template Accuracy, and (3) 92.9\% Recall Template Accuracy when using the prompt examples from the same log source of the target log message, outperforms the current best method by 6.5\%, 24.7\%, 18.8\%, respectively.
\modify{Furthermore, \methodname exhibits stable parsing quality across 16 different datasets, with standard deviations of only 3.58\%, 7.66\%, and 7.65\% for the three accuracy metrics.}

This paper makes the following main contributions:
\begin{itemize}
        \item It proposes \methodname, the first general log parsing framework that exploits the in-context learning ability for log parsing by mining common patterns in prompt examples.
        \item It designs a general prompt format to explicitly control the output structure and ensure the quality of the generated log template from LLMs, which can be generalized to enhance other prompting methods. 
        \item It presents the evaluation of \methodname on 16 public datasets using three different metrics. The results show that \methodname achieves the SOTA performance, outperforming all existing log parsing tools.
\end{itemize}

\section{Motivation And Background}

\begin{figure}[tbp]
\centering
\includegraphics[scale=0.5]{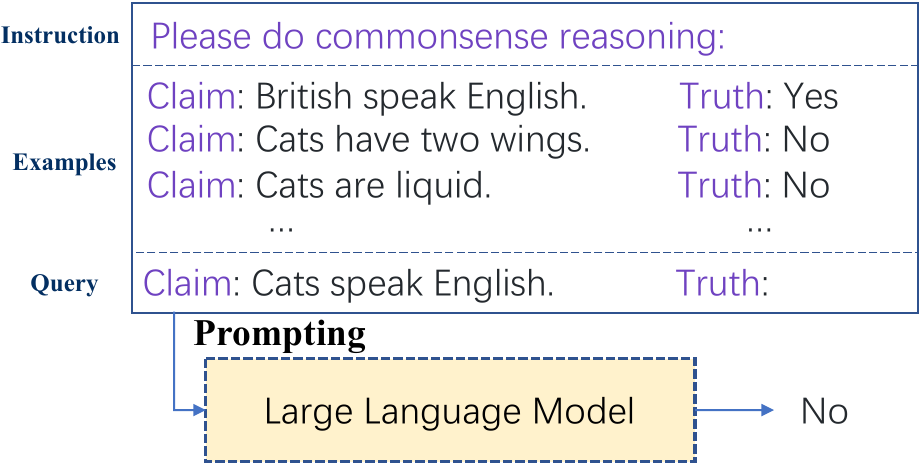}   
\caption{An example of ICL on commonsense reasoning task.}
\label{fig:iclexample}
\end{figure}


\subsection{Log Parsing}~\label{sec:parsing}

Log parsing is the initial and most critical step in automated log analysis~\cite{survey,benchmark}, which aims to extract log templates from semi-structured logs to produce structured logs.
Specifically, as shown in Fig.~\ref{fig:parsingexample}, in log parsing, a log parser is required to first extract the log headers, typically consisting of timestamps (\ie specific date and time) and verbosity level (\eg \textit{ERROR}, \textit{INFO}, and \textit{DEBUG}).
Since log headers are automatically generated by the logger, their format is fixed and can usually be extracted easily.
Therefore, log parsers primarily focus on extracting constants from log messages as log templates.
An intuitive method to extract constants from log messages is to manually design regular expressions for each log template.
However, this method is impractical for modern software systems due to the continuously increasing volume of log templates in industrial scenarios~\cite{spine,robustlog}, resulting in an unaffordable effort to maintain log parsing rules~\cite{survey}.
Therefore, numerous data-driven log parsers have been proposed, including unsupervised parsers~\cite{ael,spell,logram,drain} and supervised parsers~\cite{uniparser,semparser,logppt}.
However, they suffer from unsatisfactory effectiveness in diverse log sources due to two reasons in design as stated in Sec.~\ref{sec:intro}:
\textit{First}, unsupervised parsers utilize specially designed features or heuristics based on domain knowledge for pattern extraction, leading to failure in logs that do not match well with the features.
\modify{}
\textit{Second}, supervised parsers typically require model training on target log samples to learn features, making it open to the pitfalls of unsatisfactory generalization ability, resulting in sub-optimal performance on logs with a large sample diversity.
Therefore, we believe a properly evaluated effective log parsing methodology that is suitable for diverse logs is needed.

\subsection{Large Language Model}\label{sec:llm}


Large language models (LLMs) are deep neural networks that typically consist of 100 million to over 1 billion parameters and are obtained through self-supervised pre-training on massive valuable corpora, \eg Wikipedia, GitHub, arXiv, \textit{.etc}. 
Existing LLMs are predominantly based on the Transformer architecture. 
Depending on their specific structures, LLMs can be categorized into three main types: \textit{encoder-decoder} LLMs (\eg T5~\cite{t5}, BART~\cite{bart}), \textit{encoder-only} LLMs (\eg BERT~\cite{bert}, RoBERTa~\cite{roberta}), and \textit{decoder-only} LLMs (\eg GPT series~\cite{gpt,gpt2,gpt3}). 
These different LLM structures are often suited for various Natural Language Processing (NLP) tasks. 
For instance, encoder-only LLMs are commonly employed for tasks like text classification, while decoder-only LLMs are usually used for text generation.

The emergence of LLMs has fundamentally transformed the paradigm of NLP from training models from scratch to fine-tuning existing pre-trained models, \ie \textit{pre-train \& fine-tune} paradigm.
This shift has had profound implications, benefiting numerous tasks in NLP and other domains.
Specifically, in software engineering, fine-tuned LLMs have significant impacts on various tasks, \eg log parsing~\cite{logppt}, code review generation~\cite{auger}, fuzzing test case generation~\cite{fuzzer}, automated program repair~\cite{aprchatgpt,aprllm}, and root cause analysis~\cite{ahmed2023recommending}.
\modify{
However, this task-specific fine-tuning could excessively restrict the model to the training data, diminishing its pre-training derived generalization ability, despite improving downstream task performance.~\cite{finetuneharmgeneral}
}

\subsection{In-Context Learning}\label{sec:icl}

In recent years, with the increase in model sizes and richer training corpora, LLMs have notably grown in power. 
Researchers have discovered that, in numerous instances, merely presenting the task description as a textual prompt to pre-trained LLMs suffices for performing various tasks, eliminating the necessity for task-specific fine-tuning.~\cite{promptsurvey}
The approach is known as \textit{pre-training, prompting, and predicting}.~\cite{promptsurvey}
For example, when classifying the sentiment of the sentence “I feel good.”, we can append a prompt “I am experiencing $[MASK]$ emotion.” and ask the LLM to fill the blank $[MASK]$ with a polar word.

Furthermore, since the advent of GPT-3, researchers have observed that LLMs possess the capability of learning in the prompt context to perform inference.
This few-shot prompting is known as \textit{in-context learning} (ICL).~\cite{gpt3,icl}. 
Specifically, if a prompt contains (1) a clear \textit{instruction} specifying a particular task, (2) a few \textit{examples} with ground-truth labels providing task-specific knowledge, and (3) a \textit{query} from this particular task, the LLM can generate an answer for the query by mining the semantics between the examples and the query based on the instruction.
For instance, Fig.~\ref{fig:iclexample} illustrates an example of ICL on commonsense reasoning.
Multiple studies have demonstrated that LLMs perform well on various complex problems with ICL, such as fact retrieval~\cite{calibrate} and mathematical reasoning~\cite{chainofthought}. 
People typically believe that LLMs have gained the ability to complex reason from a large amount of pre-training data, which enables LLMs to generate the expected answer of the query based on the demonstration examples in the prompt~\cite{icl}.

Since the ICL directly uses the knowledge obtained from pre-trained LLMs along with the information in the instruction and examples in the prompt, there is no significant gap between pre-training and downstream tasks~\cite{nomoreft}, making the output more consistent with the human expectation.
It avoids the issues of over-reliance on specific data and enables high-quality predictions on the overall target log sources.
To this end, we proposed \methodname, the first attempt to automatically and efficiently accomplish the log parsing task within the framework of ICL.

\begin{figure*}[tpb]
\centering
\includegraphics[scale=0.85]{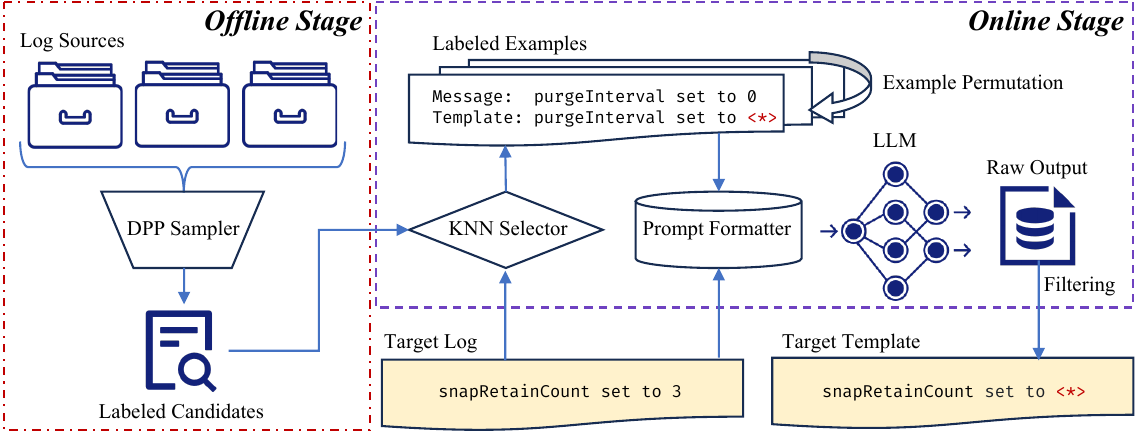}   
\caption{The workflow of \methodname framework.
The candidates need to be sampled and labeled before log parsing.
}
\label{fig:workflow}
\end{figure*}

\section{\methodname}~\label{sec:approach}

\subsection{Overveiw}

\modify{
Our method, \methodname, performs high-quality log parsing under the ICL paradigm without training.
As depicted in Fig.~\ref{fig:workflow}, \methodname is composed of an offline sampling stage and an online parsing stage.
Specifically, \methodname uses GPT-3~\cite{gpt3}, a pre-trained LLM that excels in multiple natural language tasks, as the backbone. 
Before parsing begins, \methodname samples a small number of logs from offline log data, maximizing the differences between samples, and provides them to developers for annotation (\ie supplying templates for each log) to construct a candidate set. 
During the parsing process, \methodname selects the five most similar labeled examples from the candidate set based on the semantics of the target log. 
These examples are then assembled in a specially designed prompt format and provided for in-context inference. 
Subsequently, the LLM learning semantics in the prompt context to analogy the template of the target log. 
Since \methodname does not necessitate model training or the handcrafted features in both the offline and online stages, it mitigates the potential limitations as discussed in Sec.~\ref{sec:intro}.
}

\subsection{Problem Definition}

\begin{figure}[tbp]
\centering
\includegraphics[scale=0.78]{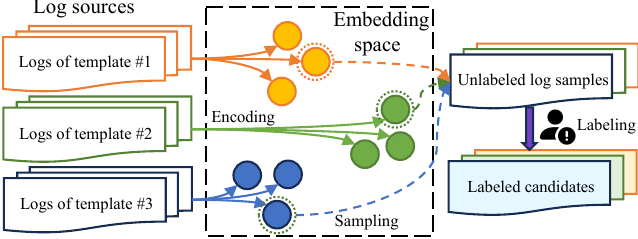}   
\caption{\modify{DPP Sampler: Sampling the most diverse logs as the candidates of the prompt examples from offline logs for subsequent online parsing.}
}
\label{fig:dpp}
\end{figure}

This paper explicitly defines the log parsing task as a log template generation task for log messages.
Specifically, for a raw log message $x$ containing $n$ tokens after tokenization (denoted as $x=[t^r_1, t^r_2, ..., t^r_n]$) as input, our proposed method, \methodname, are required to generate a sequence $y$ consisting $n$ tokens (denoted as $y=[t^p_1, t^p_2, ..., t^p_n]$) within the locator pair (defined in Sec.~\ref{sec:prompttemplate}) as the log template of $x$.
The difference between $x$ and $y$ is that all variables in $x$ are substituted by wildcard $\langle$*$\rangle$ in $y$.
The relationship $\mathcal{F}$ of $t^r_i$ and $t^p_i$ can be written as follows:

\begin{equation}
    t^p_i = \mathcal{F}(t^r_i) = 
    \left
    \{
    \begin{aligned}
    & ``\langle*\rangle" &  \ \ \text{if $t^r_i$ is variable} \\
    &  \ \ t^r_i &  \ \  \text{if $t^r_i$ is constant}
    \end{aligned}
    \right.
\end{equation}

For example, as shown in Fig.~\ref{fig:parsingexample}, for a raw log message \texttt{"Setting block size to 1919810"}, the generated log template is \texttt{"Setting block size to  $\langle$*$\rangle$"}. 
In the online parsing (inference) stage, the generated sequence will be considered as the log template, and the different tokens between the raw log and the generated sequence will be served as parameters.

\subsection{Model Backbone}~\label{sec:backbone}

\begin{figure}[tbp]
\centering
\includegraphics[scale=0.78]{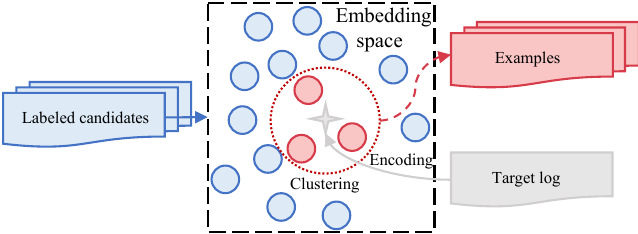}   
\caption{\modify{KNN Selector: Selecting the most similar logs along with their labels as examples from the candidate set for the subsequent prompting.}
}
\label{fig:knn}
\end{figure}

The performance of the large language model is a key factor in the success of ICL. 
Considering that log messages are semi-structured sentences that are mainly composed of natural language descriptions (\ie log template), we chose GPT-3~\cite{gpt3}, an LLM that is pre-trained on an extremely large amount of semantic information from the open-source corpus, as the backbone for \methodname. 
We will not delve into the architectural details of GPT-3 in this paper.
However, it is worth noting that from GPT-3 onwards, it was gradually discovered that LLMs emerged with the ability to perform ICL, which enables LLMs to explore new knowledge from a small number of demonstration examples and to exploit it for the query to generate an appropriate answer in the prompt~\cite{icl}. 
It is the reason that motivates us to adopt GPT-3 as our backbone, as stated in Sec.~\ref{sec:icl}.
Since \methodname utilizes the LLM in a black-box manner, the backbone can be replaced at will as long as the model or relevant API is accessible.

\subsection{\modify{Candidate Sampling}}~\label{sec:candidate}

\begin{algorithm}[tpb]
	\renewcommand{\algorithmicrequire}{\textbf{Input:}}
	\renewcommand{\algorithmicensure}{\textbf{Output:}}
	\caption{DPP: Enhanced Candidate Sampling}
    \label{alg:dpp}
	\begin{algorithmic}[1]
    \REQUIRE logs from target sources $\mathcal{X}=\{x_i\}^N_{i=1}$, encoder $\epsilon(\cdot)$ ,candidate number: $K$
    \STATE Init: $\boldsymbol{c_i}=[]$, $j = \arg \max_{i \in Z} \log(d_i^2)$, $\mathcal{C} = \mathcal{C}\cup \{x_j\}$
    \STATE $V = \epsilon(\mathcal{X})$
    \STATE $L = \text{sim}(V, V)$, $d_i^2$ = $L_{ii}$ 
    \WHILE{$\mathcal{C}\text.size \leq K$}
    \FOR {$i \in \mathcal{X}\backslash \mathcal{C}$}
    \STATE $e_i = (L_{ji} - c_j^T c_i)/d_j$
    \STATE $\boldsymbol{c_i} = \begin{bmatrix}\boldsymbol{c_i} & e_i \end{bmatrix}$
    \STATE $d_i^2 = d_i^2 - e_i^2$
    \ENDFOR
    \STATE $j = \arg \max_{i \in Z} \log(d_i^2)$, $\mathcal{C} = \mathcal{C}\cup \{x_j\}$
    \ENDWHILE
    \ENSURE Candidate set $\mathcal{C}$
	\end{algorithmic} 
\end{algorithm}




\modify{Before log parsing, \methodname needs to construct a set of labeled log samples as the candidates of prompt examples for subsequent ICL inference.
However, since the collected logs are unlabeled and manual labeling is labor-intensive, people usually only sample a subset of logs for labeling.}
To enhance the robustness of \methodname and reduce the effort of manual labeling, we use Determinantal Point Process (DPP)~\cite{dpp}, a classic algorithm for maximizing the sample diversity, to select a small number of logs as candidates from the offline data for labeling.
By explicitly optimizing the diversity of the candidate set, \methodname can achieve an equitable selection of unlabeled log samples from different templates, thereby reducing the potential risk of inductive bias that may arise from the unbalanced sample distribution. 
Moreover, using DPP for sampling yields stable results for each offline log dataset, which could mitigate the impact of randomness in the entire \methodname workflow.

The detailed algorithm is shown in Algo.~\ref{alg:dpp}.
Specifically, \methodname first encodes log samples in the provided log dataset $\mathcal{X}=\{x_i\}^N_{i=1}$ from text sequence $x_i$ to vector $v_i$ via OpenAI embedding model, and then concatenates them to an embedding matrix $V=\epsilon(\mathcal{X})=[v_1, v_2, ...,v_N]$.
Then, \methodname calculates the similarity matrix $L=\text{sim}(V, V)$ (\ie \textit{kernel matrix} in original paper~\cite{dpp}).
Based on the similarity matrix $L$, \methodname can iteratively select samples that \textit{maximize the marginal gain in 
total dissimilarity} for the current candidate set in a greedy manner until satisfies the maximum candidate set capacity $K$.
In our implementation, we use the cosine distance to evaluate the similarity between any two tensors $\boldsymbol v_i$ and $\boldsymbol v_j$, which is shown in Eq.~\ref{eq:1}
\begin{equation}\label{eq:1}
    \text{sim}(\boldsymbol v_i ,\boldsymbol v_j) := \cos(\boldsymbol v_i,\boldsymbol v_j) = \frac{\boldsymbol v_i^T \boldsymbol v_j}{\Vert \boldsymbol v_i\Vert_2\Vert \boldsymbol v_j\Vert_2},
\end{equation}
Therefore, we can obtain a stable candidate set $C$ for the following in-context inference stage.



\subsection{Example Selection}~\label{sec:promptexample}

During log parsing, \methodname required selecting several log examples in the candidate set as the prompt examples for each target log (\ie the query).
\modify{Recently, several studies~\cite{epr,kate} have pointed out that the selection of examples greatly influences the performance of ICL-based approaches. 
For instance, \textit{Liu et al.}~\cite{kate} found that using examples closer to the query can significantly enhance the quality of inference.}
To strengthen the accuracy of \methodname, we adopt $k$-Nearest Neighbor ($k$NN), a straightforward clustering algorithm, to select several log examples that are most similar to the query as demonstration examples.
Through deliberate selection of log samples that resemble the query as examples, \methodname effectively learns the semantics and common patterns of the query from data that aligns closely with its characteristics.
Furthermore, the examples chosen from $k$NN exhibit higher similarity to one another, leading to a more compact data distribution in the prompt. This facilitates the LLM in comprehending the log semantics within the context.




The detailed algorithm is shown in Algo.~\ref{alg:knn}
Specifically, \methodname also begins by encoding all log candidates $x_i \in \mathcal{C}$ as well as the query $x_q$ into embedded vectors $v_i$ and $v_q$. 
Then, for each vectorized query $v_q$, it calculates the similarity sim$(v_q, v_i)$ between it and all candidates $v_i$.
After iterating over the entire candidate set, \methodname generates a distance map $\mathcal{D}$ to record the similarity between the query vector $v_q$ and all candidate queries $v_i$. 
Subsequently, by querying the distance map $\mathcal{D}$, the top-$k$ ($k$=5 by default) most similar log samples can be extracted as demonstration examples $\mathcal{E}$.
In our implementation, the similarity metric sim$(v_i, v_j)$ also represents the cosine distance as shown in Eq.~\ref{eq:1}.



\subsection{Example Permutation}~\label{sec:promptpermutation}



To further strengthen the parsing performance, \methodname deliberately reorders the selected examples $\mathcal{E}$ in the \textit{ascending order} based on their similarity to the query $\text{sim}(v_q,v_i)$ during prompt construction.
This is because recent research~\cite{calibrate} suggests that LLM's ICL exhibits a \textit{recency bias} characteristic, whereby the LLM is more susceptible to the inductive bias of the examples closer to the query when generating an answer prediction. 
By ordering examples based on similarity, the example closer to the query is more similar to the query, making it easier for the LLM to learn the hidden relations between the last example and its label, and generate a prediction closer to the query's ground-truth label. 
The advantages of this approach will be further elaborated in subsequent experimental results~\ref{sec:rq2}.

\begin{algorithm}[tpb]
	\renewcommand{\algorithmicrequire}{\textbf{Input:}}
	\renewcommand{\algorithmicensure}{\textbf{Output:}}
	\caption{$k$NN: Augmented Example Selection}
    \label{alg:knn}
	\begin{algorithmic}[1]
    \REQUIRE query $x_q$, candidate set $\mathcal{C} = \{x_i\}^K_{i=1}$, encoder $\epsilon(\cdot)$, example number:$k$
    \STATE Init: distance map $\mathcal{D}=\emptyset$ and $\mathcal{E}=\emptyset$
    \STATE $v_q = \epsilon(x_q)$
    \FOR {$x_i \in \mathcal{C}$}
    \STATE $v_i = \epsilon(x_i)$
    \STATE $d_{qi} = \text{sim}(v_q, v_i)$
    \STATE $\mathcal{D} = \mathcal{D}\cup \{d_{qi}: x_i\}$
    \ENDFOR
    \STATE extract values with top-$k$ largest keys from $\mathcal{D}$ to $\mathcal{E}$
    \ENSURE Example set $\mathcal{E}$
	\end{algorithmic} 
\end{algorithm}

\subsection{Prompt Format}~\label{sec:prompttemplate}

\begin{figure}[tbp]
\centering
\includegraphics[scale=0.55]{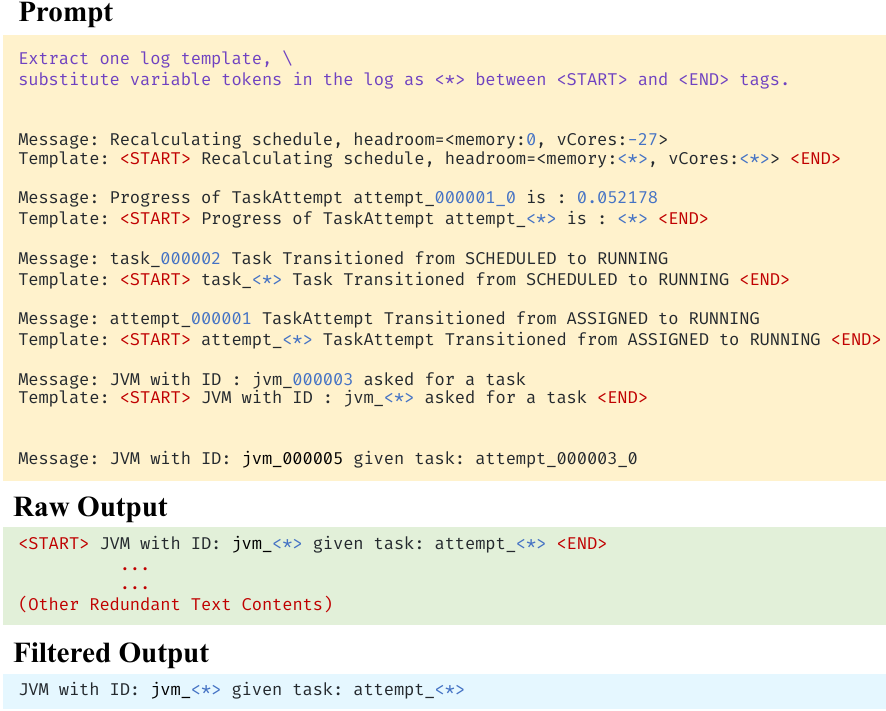}   
\caption{An example of a prompt and its expected output.
}
\label{fig:prompttemplate}
\end{figure}

A carefully designed prompt format that consists of a clear and concise instruction along with several relevant examples are necessary for LLM to comprehend the log parsing task, as discussed in Sec.~\ref{sec:icl}.
However, providing a naive prompt without output restriction could be impractical for log parsing since the LLM may produce extraneous text beyond the query answer, potentially compromising parsing quality. 
For instance, in the sentiment classification task depicted in Fig.\ref{fig:iclexample}, the output might include explanatory content, such as ``It is negative. Because awful is not a positive word", whereas only the word ``negative" is relevant to the task. 
Thus, even if the anticipated response is present in the generated text, automatically extracting the template precisely is still challenging.

\modify{To filter the redundant contents in the raw output}, we include a locator pair $``\langle$START$\rangle"$ and $``\langle$END$\rangle"$ in the prompt to limit the answer we require within the locators. 
Specifically, in \methodname, each example's log template is enclosed by the locator pair.
With the guide of the restricted format for demonstration, LLM is inclined to generate the log template within the same locator pair via analogy~\cite{icl}.
An example of a complete prompt is presented in Fig.\ref{fig:prompttemplate}, where the last example is the most similar one to the query, and the expected raw output is \texttt{"$\langle$START$\rangle$ JVM with ID: jvm\_$\langle$*$\rangle$ given task: attempt\_$\langle$*$\rangle$ $\langle$END$\rangle$"} \modify{along with other redundant output generated by the model}.
Then, \methodname can simply utilize regular expressions to extract the expected log template in the output. 
Notably, even with locators, the output often generates redundant text after the expected answer.
However, we can easily extract the expected answer with the help of locators and just simply ignore the rest of redundant text.


\section{Experiment Setup}~\label{sec:design}

\subsection{Research Question Design}
We conducted extensive experiments on 16 public datasets to answer the following research questions:

\begin{itemize}
    \item \modify{RQ1: How effective and stable is \methodname?}
    \item \modify{RQ2: How do different settings affect \methodname?}
\end{itemize}

Specifically, RQ1 investigates \methodname's parsing effectiveness (\ie accuracy and robustness~\cite{benchmark,logppt}) on various log sources through comparison to other cutting-edge log parsers.
RQ2 further discusses the contribution of each component of \methodname through ablation study.
Furthermore, it also explores the impact of different configurations through comparative experiments.

\subsection{Environment and Implementation}

In our experiments, we utilized HTTP requests to invoke the OpenAI APIs and interact with the LLM (\ie GPT-3 \texttt{curie}) and encoder (\ie \texttt{text-search-babbage-query-001}) to get the raw output response of the origin prompt.
We also employed Python 3.9 to implement Algo.~\ref{alg:dpp}, Algo.~\ref{alg:knn}, and various evaluation scripts to construct the complete prompt construction in a local machine with Ubuntu 20.04.5 LTS.
\modify{Specifically, in Algo.~\ref{alg:dpp}, \methodname samples 200 logs from each log dataset to construct the candidate set, while in Algo.~\ref{alg:knn}, it selects 5 labeled examples from the candidate set to create prompt for each query.
These settings remain consistent throughout the evaluation process.}

\subsection{Datasets}

Our experiments are conducted on 16 public log datasets from LogPAI \cite{benchmark}, the most widely-used benchmark in log analysis.
The LogPAI datasets cover a variety of log types, including distributed systems, standalone software, supercomputers, PC operating systems, mobile systems, and microservices.
In each dataset, all logs are labeled with ground-truth templates in advance with a unique Event ID.
The logPAI datasets have been widely used in multiple log parsers' evaluations, and we will keep following them to use LogPAI datasets as standard benchmarks.

\subsection{Metrics}

Following recent studies~\cite{guideline,logppt}, our evaluation uses three metrics: Parsing Accuracy (PA), Precision Template Accuracy (PTA), and Recall Template Accuracy (RTA), where the last two metrics are also known as Template Accuracy (TA)~\cite{guideline}.
Specially, we use PA to evaluate the parsing performance at the message level and use PTA and RTA to evaluate the parsing performance at the template level.
The concepts of these metrics are as follows:

\subsubsection{Parsing Accuracy (PA)}
Parsing Accuracy (PA), also known as Message Level Accuracy (MLA)~\cite{uniparser}, is defined as the ratio of "correctly parsed" log messages over the total number of log messages, where a log is considered to be "correctly parsed" if and only if all constants and variables are exactly divided in the log template.
Since PA is not related to the number of templates, but only to the total number of parsed log messages, we use PA as a message-level evaluation metric to assess the most basic parsing capability of the parsers.

\subsubsection{Template Accuracy (TA)}
Template Accuracy (TA) is proposed by Kan~\etal~\cite{guideline}, which gives template-level guidance of parsing quality.
Specifically, TA comprises two metrics, Precision Template Accuracy (PTA) and Recall Template Accuracy (RTA). 
PTA measures the ratio of correctly identified templates to the total number of identified templates, while RTA measures the ratio of correctly identified templates to the total number of ground-truth templates. 
The concept of "correctly identified" templates means that all log messages from this log template are "correctly parsed" (defined in the PA context).
Therefore, it is obvious that PTA and RTA are more stringent metrics than PA, and we aim to use them to showcase the strong parsing ability of \methodname.

Additionally, we do not continue to adopt Grouping Accuracy (GA), which is widely used for unsupervised parser evaluation in the past.
This is mainly due to two reasons. 
\textit{Firstly}, GA has a broader definition of "correct parsing," which reduces its reference value for supervised parsers.
Specifically, GA only requires that logs belonging to the same template be correctly clustered, without considering whether constants and variables in the logs are distinguished correctly. 
\textit{Secondly}, although GA aims to evaluate the clustering ability of logs at the \textit{template} level, its value is affected by the \textit{message} number in each cluster, rather than the total number of templates~\cite{guideline}. 
Compared to GA, TA can provide more effective evaluation capabilities at the template level.~\footnote{We also provide \textbf{GA} evaluation in anonymous repository as a reference.}

\subsection{Baselines}\label{sec:baseline}

We selected LenMa~\cite{lenma}, Spell~\cite{spell}, Drain~\cite{drain}, and Logram~\cite{logram}, the top-performing unsupervised log parsers, and LogPPT~\cite{logppt}, the current state-of-the-art supervised log parser, as our baselines for comparison.
Specifically, these unsupervised log parsers are based on handcrafted features, \ie log length, longest common subsequence, prefix tree, and n-gram, while LogPPT fine-tunes a pre-trained language model, RoBERTa~\cite{roberta}, and adapts it to log parsing.
Notably, supervised parsers typically perform better than unsupervised parsers, and the only parser that \methodname can compare in an \textit{apple-to-apple} manner is LogPPT.
\modify{
However, as we cannot reproduce the results of LogPPT from its implementation on GitHub~\cite{logpptrepo}, we solely rely on the reported paper results for comparison. 
Specifically, in the GitHub implementation, LogPPT employs the Adaptive Random Testing algorithm to select 32 labeled logs for model training and evaluates on all 2,000 logs in each dataset. 
Notably, LogPPT's accuracy saturates after using 32 logs as \textit{supervision} for training (see Fig.~5 in the LogPPT paper~\cite{logppt}), while \methodname only uses 5 examples selected from the DPP sampled 200 candidates as \textit{supervision} for online parsing \textit{without training}.
Thus, we believe the comparison is fair.
}
The baselines' results are consistent with the reported data in previous research.~\cite{logppt}

\section{Evaluation Results}~\label{sec:eval}

\begin{table*}[tbp]
\caption{Accuracy comparison with the cutting-edge log parsers on LogPAI datasets (\%).}
\label{table:acc}
\setlength\tabcolsep{3.5pt}
\centering
\scalebox{1}{
\begin{tabular}{c|ccc|ccc|ccc|ccc|ccc|ccc}
    \toprule
    \multirow{2}*{\textbf{Dataset}} & \multicolumn{3}{|c|}{\textbf{LenMa}} & \multicolumn{3}{|c|}{\textbf{Spell}} & \multicolumn{3}{|c|}{\textbf{\textbf{Drain}}} & \multicolumn{3}{|c|}{\textbf{Logram}} & \multicolumn{3}{|c|}{\textbf{LogPPT}} & \multicolumn{3}{|c}{\textbf{\methodname}} \\
    & PA & PTA & RTA & PA & PTA & RTA & PA & PTA & RTA & PA & PTA & RTA & PA & PTA & RTA & PA & PTA & RTA \\
    \midrule
    Andriod&	72.2&	59.0&	60.1&	24.1&	24.4&	27.2&	73.0&	56.6&	62.0&	42.8&	38.7&	27.2&	76.7&	58.4&	68.4&	\textbf{94.3}&	\textbf{84.6}&	\textbf{86.1}\\
    Apache&	29.3&	42.9&	50.0&	28.5&	25.0&	16.7&	\textbf{100.0}&	\textbf{100.0}&	\textbf{100.0}&	50.9&	33.3&	83.3&	99.4&	83.3&	83.3&	\textbf{100.0}&	\textbf{100.0}&	\textbf{100.0}\\
    BGL&	15.4&	9.6&	25.8&	32.9&	12.1&	12.5&	44.4&	33.9&	30.8&	17.0&	15.7&	16.7&	97.0&	68.6&	78.3&	\textbf{98.7}&	\textbf{91.7}&	\textbf{92.5}\\
    Hadoop&	24.2&	20.6&	28.9&	19.6&	20.8&	23.7&	43.9&	36.8&	34.2&	37.0&	18.7&	17.5&	89.5&	54.0&	58.8&	\textbf{99.6}&	\textbf{98.3}&	\textbf{99.1}\\
    HDFS&	12.5&	37.5&	42.9&	48.7&	57.1&	57.1&	95.9&	81.3&	92.9&	1.8&	19.4&	42.9&	90.2&	85.7&	85.7&	\textbf{99.9}&	\textbf{86.7}&	\textbf{92.9}\\
    HealthApp&	28.9&	3.4&	46.7&	15.2&	21.5&	18.7&	24.1&	8.3&	34.7&	26.3&	3.5&	25.3&	78.9&	85.3&	85.3&	\textbf{99.6}&	\textbf{94.7}&	\textbf{96.0}\\
    HPC&	67.1&	12.9&	50.0&	53.2&	38.3&	39.1&	67.2&	38.8&	41.3&	67.9&	37.7&	43.5&	94.7&	73.6&	84.8&	\textbf{99.7}&	\textbf{91.5}&	\textbf{93.5}\\
    Linux&	13.2&	41.0&	41.4&	10.9&	17.3&	14.7&	19.4&	43.4&	42.2&	18.5&	13.5&	4.3&	94.9&	47.5&	49.1&	\textbf{99.7}&	\textbf{95.8}&	\textbf{95.8}\\
    Mac&	15.5&	14.6&	19.4&	3.3&	5.7&	5.9&	27.7&	21.2&	24.9&	25.2&	16.4&	20.2&	67.3&	43.6&	53.4&	\textbf{86.3}&	\textbf{67.6}&	\textbf{66.6}\\
    OpenSSH&	15.5&	25.0&	23.1&	12.7&	25.0&	23.1&	53.4&	52.0&	50.0&	48.2&	29.0&	34.6&	\textbf{97.6}&	48.9&	84.6&	93.9&	\textbf{96.2}&	\textbf{92.6}\\
    OpenStack&	19.1&	18.7&	60.5&	0.0&	0.0&	0.0&	18.7&	5.5&	39.5&	11.2&	9.3&	9.3&	90.7&	84.4&	88.4&	\textbf{99.9}&	\textbf{95.3}&	\textbf{95.3}\\
    Proxifier&	50.6&	10.6&	87.5&	47.8&	22.2&	25.0&	52.7&	26.9&	87.5&	0.0&	0.0&	0.0&	\textbf{100.0}&	\textbf{100.0}&	\textbf{100.0}&	\textbf{100.0}&	\textbf{100.0}&	\textbf{100.0}\\
    Spark&	2.3&	12.4&	33.3&	33.6&	34.5&	27.8&	37.6&	50.0&	41.7&	27.5&	3.1&	22.2&	99.1&	60.0&	58.3&	\textbf{99.9}&	\textbf{97.2}&	\textbf{97.2}\\
    Thunderbird&	17.1&	27.1&	34.2&	3.9&	19.1&	16.8&	19.1&	29.9&	36.9&	12.8&	20.8&	6.7&	92.6&	50.6&	59.1&	\textbf{98.5}&	\textbf{91.8}&	\textbf{95.3}\\
    Windows&	26.6&	37.9&	44.0&	0.4&	11.5&	12.0&	69.6&	46.3&	50.0&	37.4&	15.2&	10.0&	98.3&	55.4&	72.0&	\textbf{99.8}&	\textbf{91.8}&	\textbf{90.0}\\
    Zookeeper&	45.7&	30.0&	36.0&	45.3&	22.0&	26.0&	49.8&	39.1&	36.0&	51.6&	26.1&	24.0&	99.0&	74.1&	86.0&	\textbf{99.8}&	\textbf{88.7}&	\textbf{94.0}\\
    \hline
    Average& 28.5&	25.2&	42.7&	23.8&	22.3&	21.6&	49.8&	41.9&	50.3&	29.8&	18.8&	24.2&	91.6&	67.4&	74.1&	\textbf{98.1}&	\textbf{92.1}&	\textbf{92.9}\\
    \bottomrule
\end{tabular}
}
\end{table*}

\subsection{RQ1: How effective and stable is \methodname?}~\label{sec:rq1}

This section compares the accuracy and robustness of \methodname with other state-of-the-art log parsers on 16 publicly available log datasets. 
The study evaluates Parsing Accuracy (PA), Precision Template Accuracy (PTA), and Recall Template Accuracy (RTA) to demonstrate the accuracy of the parsers. 
Following the previous research~\cite{benchmark,logppt}, the accuracy distribution across the 16 datasets is also reported to assess the methods' stability and robustness.


\modify{
Notably, log data might contain multiple completely identical log instances. 
If one such log is included in the candidate set, it could result in data leakage when parsing other identical logs, as these identical logs might serve as both the prompt example and the query simultaneously (even though they are not exactly the "same" log).
To ensure fair evaluation, in this experiment, we ask \methodname to avoid selecting example logs that are identical to the query. 
It continues selecting examples from the candidate set in order of similarity until no identical candidate is included in the examples. 
This also enables \methodname to be directly evaluated on all 2,000 logs in each dataset without data leakage in the prompt examples, ensuring the same evaluation data used for other unsupervised parsers and LogPPT, as discussed in~Sec.\ref{sec:baseline}. 
Note the accuracy of \methodname in practice could be higher because it is common to observe identical logs during the continuous log collection process.
}

\subsubsection{Accuracy}

Accuracy is the most critical characteristic to evaluate the effectiveness of a log parser~\cite{benchmark}.
Any minor parsing errors can lead to a significant impact on the performance of downstream tasks of log analysis~\cite{evaluationstudy}.
In this part, we focus on evaluating the accuracy metrics of the log parser on each dataset, as well as its overall average accuracy metrics. 
These metrics provide a distinct representation of the global strengths and weaknesses of each parser for each metric, as well as the relative strengths and weaknesses of each parser on each dataset.

The results are shown in Table.~\ref{table:acc}.
To visualize the advantage of \methodname, we marked the best results in each evaluation metric on each dataset, as well as their average value on all datasets in bold font.
It is clear that \methodname demonstrates state-of-the-art Parsing Accuracy (PA), Precision Template Accuracy (PTA), and Recall Template Accuracy (RTA) on all 16 LogPAI log datasets, outperforming the best DL-based parser, LogPPT~\cite{logppt}, which is proposed recently. 
Specifically, regarding the widely used PA metric, \methodname achieved an average PA of 98.1\% across 16 datasets, with 100\% PA on 4 datasets and PA exceeding 95\% on 13 datasets \modify{(including 2 datasets with 100\% PA)}, and no dataset had a PA below 80\%. 
This outperforms any previous log parser by far. 
For instance, \methodname outperformed the current best log parser LogPPT on 14 out of 16 datasets in terms of PA, with less than 4\% PA gap on the only dataset where it was not superior, and another dataset tied at 100\% accuracy. 
Additionally, \methodname's advantage is more evident in PTA and RTA metrics. 
It achieved an average PTA and RTA of 92.1\% and 92.9\%, across 16 datasets, with 12 and 14 datasets having PTA and RTA metrics not lower than 90\%, respectively.
This performance significantly exceeds that of LogPPT by 21.0\% and 16.5\%, respectively, where LogPPT only achieved an average PTA and RTA of 67.4\% and 74.1\%. 
These findings demonstrate the significant accuracy advantage of \methodname over current parsers.

Furthermore, \methodname's accuracy advantage over other traditional log parsers is even more apparent. 
For instance, compared to the strongest traditional log parser, Drain, \methodname exceeded its average PA, PTA, and RTA by 48.3\%, 50.2\%, and 42.6\%, respectively. 
We believe that this phenomenon is due to the fact that traditional log parsers, which are based on expert-designed heuristic rules or handcrafted features, struggle to correctly distinguish between variables and constants in different types of log data, and can only coarsely cluster log messages with the same template. 
Therefore, to fairly demonstrate \methodname's accuracy, we primarily compared it with the current SOTA supervised log parser, LogPPT.

\subsubsection{Robustness}

\begin{figure*}[tpb]
\centering
\includegraphics[scale=0.42]{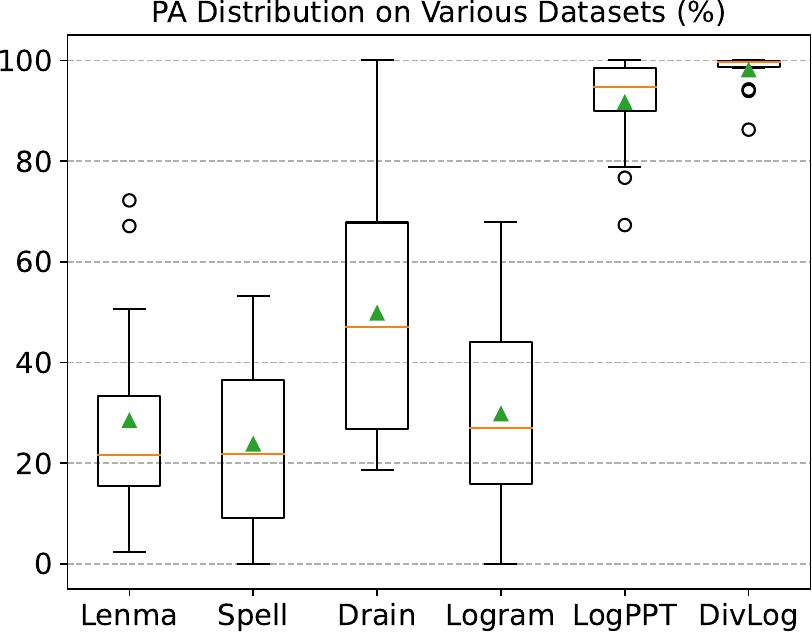}   
\includegraphics[scale=0.42]{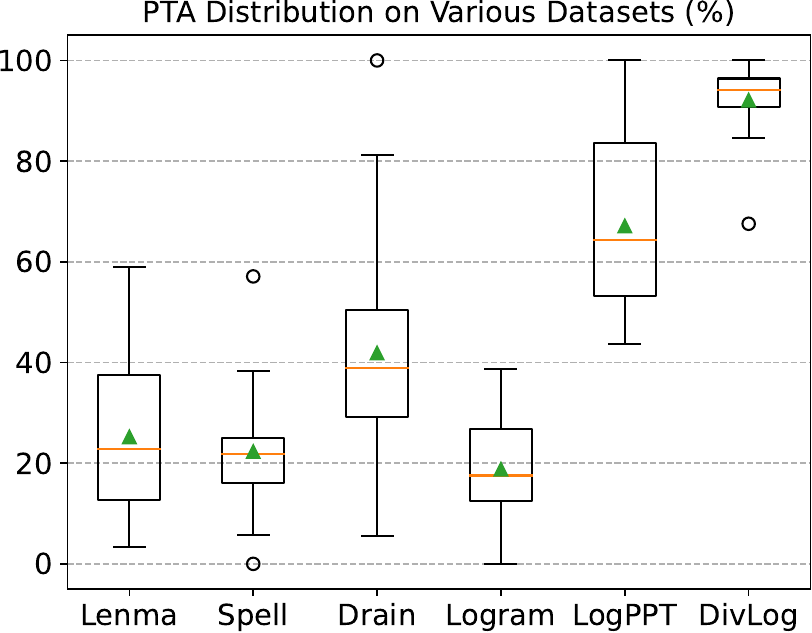}   
\includegraphics[scale=0.42]{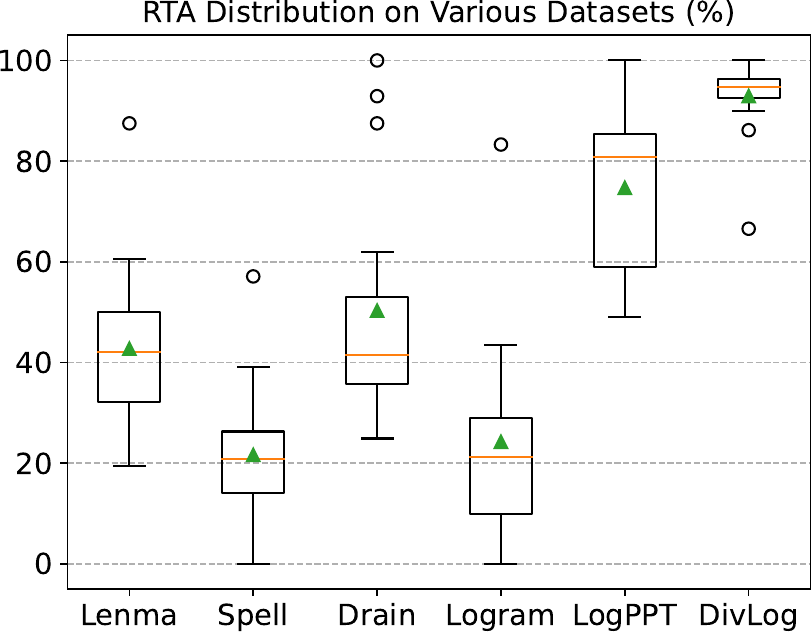}   
\caption{\modify{Accuracy on different datasets (\%). Our approach, DivLog, achieves the most stable performance among 16 datasets.}}
\label{fig:distribution}
\end{figure*}

Robustness is an important characteristic to measure the versatility of a log parser~\cite{benchmark}.
A practical log parser should perform consistently on logs generated in multiple production environments~\cite{benchmark}.
In this part, we focus on evaluating the robustness of each log parser through the analysis of the distribution of metrics among 16 log datasets.
By analyzing the distribution of metrics for each log parser, we can identify the stability of their parsing ability on different datasets.

The results are shown in Fig.~\ref{fig:distribution}.
The results demonstrate that \methodname outperforms existing log parsers by exhibiting the narrowest accuracy distribution among 16 datasets on three metrics.
This indicates that \methodname possesses strong robustness across various datasets. 
\modify{Specifically, among all 16 datasets, \methodname has a standard deviation of only 3.58\%, 7.66\%, and 7.65\% on the three metrics, outperforming the currently most robust log parser, LogPPT, which has standard deviations of 9.2\%, 16.56\%, and 14.57\%. }
In addition, compared to traditional unsupervised log parsers, \methodname has a more pronounced advantage. 
We believe this advantage of \methodname stems from two reasons: 
(1) \methodname learns features of different log sources from relative log examples in prompts and generates the expected template via these features, without the need for manual feature selection or heuristic design. 
(2) Since \methodname does not train or fine-tune a model but rather leverages LLM's ICL capabilities, it is not subject to the training set, which can reduce model performance stability on various log datasets.

\subsection{RQ2: How do different settings affect \methodname?}~\label{sec:rq2}

\begin{table}[tbp]
\caption{Ablation study of components on Mac dataset (\%).}
\label{table:ablation}
\centering
\setlength\tabcolsep{3pt}
\scalebox{1}{
\begin{tabular}{lccc}
    \toprule
    ~& \textbf{PA} & \textbf{PTA} &	\textbf{RTA}\\
    \midrule
    $\text{Full}_\text{\textsc{\methodname}}$ & 86.3 & 67.6& 66.6\\
    $\text{w/o}_\text{\textsc{DPP Sampling}}$ & $75.3_{(\downarrow 11.0)}$ & $39.4_{(\downarrow 28.2)}$ & $41.3_{(\downarrow 24.9)}$ \\
    $\text{w/o}_\text{\textsc{$k$NN Comparison}}$ & $17.0_{(\downarrow 69.3)}$ & $2.3_{(\downarrow 65.3)}$ & $7.3_{(\downarrow 59.3)}$ \\
    $\text{w/o}_\text{\textsc{Locators}}$ & $70.3_{(\downarrow 16.0)}$ & $50.5_{(\downarrow 17.1)}$ & $31.7_{(\downarrow 34.9)}$ \\
    \bottomrule
\end{tabular}
}
\end{table}

This section conducts ablation experiments to discuss the contributions of each key component and the impact of different configurations in \methodname. 
Specifically, we use the Mac dataset as the benchmark for the following experiments, as its complex and diverse data (as illustrated in Sec.\ref{sec:intro}) can better illustrate the contributions of different settings to log parsing, compared to other simpler datasets.


\subsubsection{Components}

\begin{figure}[tpb]
\centering
\includegraphics[scale=0.48]{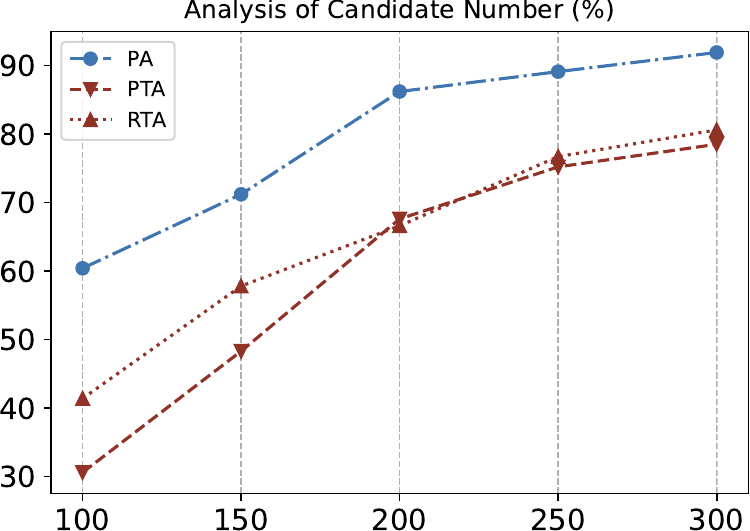}   
\caption{\modify{Analysis of candidate number on Mac dataset (\%).}}
\label{fig:cn}
\end{figure}

\begin{figure}[tpb]
\centering
\includegraphics[scale=0.48]{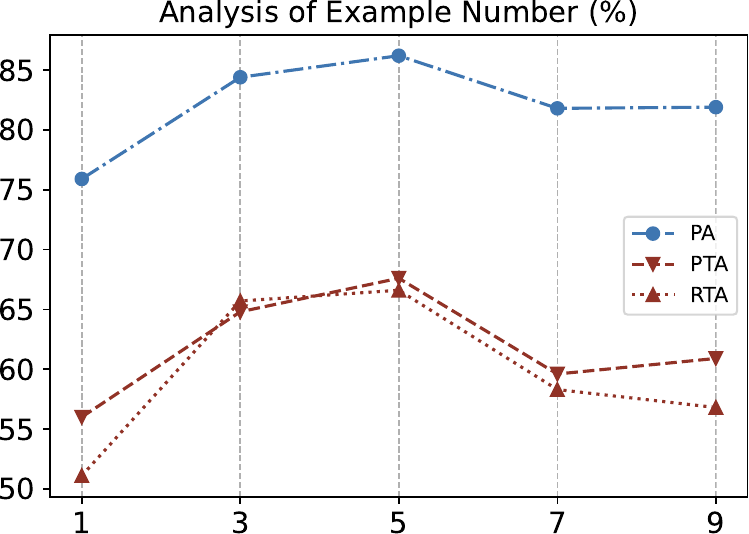}   
\caption{Analysis of example number on Mac dataset (\%).}
\label{fig:en}
\end{figure}

As shown in Fig.~\ref{fig:workflow}, \methodname comprises three main components: DPP sampling, $k$NN comparison, and prompt locators.
To evaluate the individual contributions of these components to \methodname's log parsing capability, we removed each component sequentially and measured its necessity and importance by observing the extent of the resulting accuracy drop.
Specifically, we (1) replaced the DPP algorithm with random sampling to construct the candidate set, (2) replaced the $k$NN algorithm with random selection to choose examples, and (3) removed the locators in the prompt, relying solely on the first-line output of LLM as the log template prediction.
To mitigate the impact of random bias, we repeated each experimental configuration five times and calculated their mean as the final result.

The results are shown in Table.~\ref{table:ablation}.
It is clear that removing DPP sampling, $k$NN selection, or prompt locators from \methodname leads to a significant decrease in all three accuracy metrics.
For instance, without DPP sampling or prompt locators, \methodname achieves an 11.1\% and 20.4\% lower PA than the full version, respectively.
The reason for this drop is that DPP sampling enhances the diversity of the candidate set, providing more related examples to assist in generating the log template, while prompt locators help to extract the expected log template from the raw output.
Notably, $k$NN selection plays the most crucial role in \methodname; without it, \methodname can hardly continue log parsing, achieving only 16.8\% PA, 1.8\% PTA, and 6.8\% RTA.
This is because there are enormous dissimilarities between each template and the scarcity of log samples in each template in the Mac dataset.
Without similar log messages in prompt context, LLMs can hardly mine valid semantics and common patterns among irrelevant log examples through analogy.
Thus, we conclude that these components are indispensable for the effectiveness of \methodname.

\subsubsection{Configurations}

In addition to the above-mentioned three key components, there are several configurations that significantly affect \methodname's parsing performance, \ie (1) the number of examples, (2) the permutation method, and (3) the model backbone. 
Specifically, we first analyze the impact of \methodname's performance with different numbers of examples ranging from 1 to 9.
Then, we compare the effectiveness of different permutation methods, \ie ascending, descending, and random order. 
Finally, we discuss how different model backbones in GPT-3~\cite{gpt3} series (\ie Ada, Babbage, and Curie) affect parsing performance. 
We also repeat each experiment five times and report their mean as the final result.

\textbf{Candidate Number:}
\modify{As shown in Fig.~\ref{fig:cn}, all metrics improves with an increased number of candidates.
The reason is that the sample space for examples available to each query expands, making the potential examples more suitable for the query. 
However, after reaching 200 candidates, the improvement in performance metrics slows down significantly. 
Given the considerable increase in manual labeling efforts with more candidates in practice, we limit the use to 200 candidates.}

\textbf{Example Number:}
As shown in Fig.~\ref{fig:en}, all metrics show an increasing trend when the number of examples is less than 5.
However, when the number exceeds 5, all metrics gradually decrease. 
The reason for this is that too few examples do not provide enough semantics for LLM to learn log patterns, and too many examples include irrelevant logs that do not provide effective guidance for parsing the target log message. 
Thus, in our experiments, 5 examples are sufficient, and more examples do not help significant improvement.

\begin{figure}[tpb]
\centering
\includegraphics[scale=0.48]{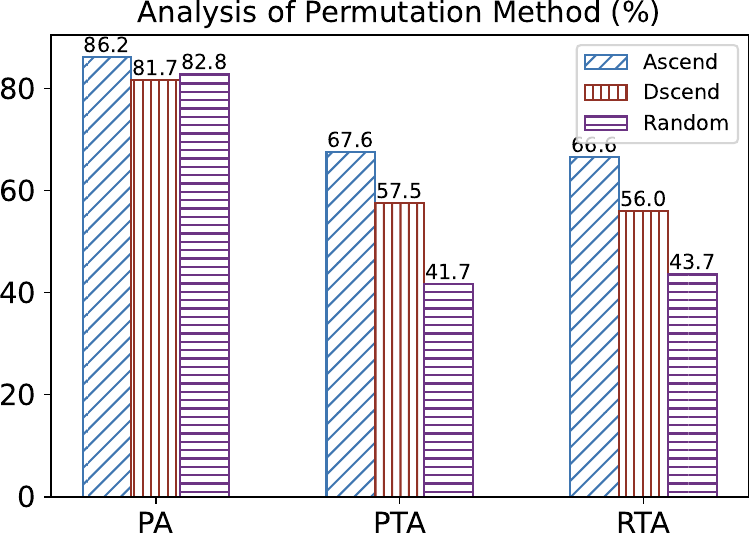}   
\caption{Analysis of permutation on Mac dataset (\%).}
\label{fig:pm}
\end{figure}

\textbf{Example Permutation:}
As shown in Fig.~\ref{fig:pm}, using ascending order performs better on all metrics than descending order and random order. 
This is consistent with the analysis in Sec.~\ref{sec:promptexample}, which suggests that due to \textit{recency bias}~\cite{calibrate}, LLMs are more likely to learn from examples that are closer to the query. 
Hence, if the most recent examples have patterns similar to the query, it enhances parsing accuracy. 
Notably, random order sometimes coincidentally creates an ascending permutation, resulting in better message-level metrics than descending order on average.
However, due to its unstable randomness, incorrectly parsed log messages may be evenly distributed among multiple templates, leading to inferior template-level performance. 
Conversely, descending order's stable ordering may not significantly affect the parsing ability on logs belonging to some templates, resulting in slightly higher PTA and RTA.

\begin{table}[tbp]
\caption{Analysis of model backbone on Mac dataset (\%).}
\label{table:backbone}
\centering
\setlength\tabcolsep{3pt}
\scalebox{1}{
\begin{tabular}{lccc}
    \toprule
    ~& \textbf{PA} &	\textbf{PTA} &	\textbf{RTA}\\
    \midrule
    Curie &86.3 &67.6 &66.6\\
    Babbage &$78.5_{(\downarrow 7.8)}$ &$58.3_{(\downarrow 9.2)}$ &$65.7_{(\downarrow 0.9)}$ \\
    Ada &$74.1_{(\downarrow 12.2)}$ & $43.6_{(\downarrow 24.0)}$ &$36.1_{(\downarrow 30.5)}$ \\
    \bottomrule
\end{tabular}
}
\end{table}

\textbf{Model Backbone:}
As shown in Table.\ref{table:backbone}, the accuracy increases with the LLM scale in \methodname framework. 
Specifically, GPT-3 consists of four different versions: Ada-350M, Babbage-3B, Curie-13B, and Davinci-175B. 
Due to limited research resources, we were only able to test \methodname's performance with the first three versions as backbones. 
Nevertheless, we believe that with the emergence of more powerful LLMs, \methodname's parsing quality will further excel.





\section{Discussion}


\subsection{Efficiency and Practicality}

Efficiency is also an aspect of evaluating the performance of log parsers.
Typically, researchers measure a log parser's efficiency by analyzing the execution time required for finishing parsing.
However, in this paper, we adopt OpenAI's API to interact with LLM and the encoder, resulting in an unmeasurable efficiency due to network latency (\ie the parsing time for each log message is almost exactly twice the network latency).
Thus, we can only qualitatively discuss potential efficiency issues in practice.

Specifically, \methodname's time cost consists of (1) constructing a candidate set, (2) selecting prompt examples, (3) encoding logs, and (4) LLM inference. 
The cost of (1) and (2) mainly comes from the DPP and the $k$NN, which are efficient and classic algorithms. 
The time complexities are $O(K^2N)$ and $O(KM)$ respectively, where $K$, $N$, and $M$ are the log scale of (i) candidate set $\mathcal{C}$, (ii) offline logs from target source $\mathcal{X}$, and (iii) incoming logs in online parsing stage.
The cost of (3) and (4) mainly depends on the model inference speed. 
\modify{For large enterprises and organizations that may face large volumes of log data, increasing the parallel computing resources might be a potential solution to mitigate the efficiency threat in LLM inference. 
Furthermore, in industrial scenarios, \methodname only needs to collect the newly produced logs as candidates for ICL inference.
It does not require additional resource consumption for retraining the models on these new log data.}
\modify{Therefore, we believe that the \methodname framework can be practical with specialized development.}


\subsection{Threat to Validity}

We identified the following major threats to validity:

\begin{itemize}
    \item \modify{\textbf{Data Leakage:} As discussed in Sec.~\ref{sec:baseline}, there is a possibility of data leakage in the prompt example due to the presence of multiple completely identical log instances. To eliminate this threat and ensure fair evaluation, in our experiments, we ask \methodname to avoid selecting example logs that are identical to the query. It continues selecting examples from the candidate set based on their similarity until no identical candidate is included, enabling direct evaluation on all 2000 logs in each dataset, consistent with the evaluation data used for other baselines.}
    \modify{In addition, due to the opacity of GPT-3's training data, it is unclear whether it was exposed to the log parsing task during pre-training, potentially leading to data leakage. To investigate its impact, we conducted zero-shot tests by removing all the examples in the prompts before evaluation, observing GPT-3 producing only invalid output and being unable to parse the templates. Therefore, we consider this threat negligible.}
    \item \textbf{Randomness:} Randomness may affect the performance through two aspects: (1) the randomness in LLMs and (2) the randomness introduced during the experiments in Sec.~\ref{sec:rq2}, including a random sampling of candidates, selection of examples, and permutation of examples. To mitigate the former threat, we reduced the randomness of the LLM by configuring it to generate consistent outputs for the same input text (\ie initialize \texttt{temperature=0}). To mitigate the latter threat, we conducted five independent experiments for each experimental setting and used the mean of the results as the final outcome.
\end{itemize}


\section{Related Works}~\label{sec:related}

Log parsing methods can be categorized into unsupervised and supervised methods according to the parsing algorithms. 

\textbf{Unsupervised log parsers}
Unsupervised log parsers leverage manually designed features to extract log templates, which have been widely explored in the past.
There are three main categories of unsupervised log parsers: frequent pattern mining-based, clustering-based, and heuristic-based methods. 
(1) \textit{Frequent pattern mining-based} methods regard the mined frequent patterns (\eg n-grams) as log templates.
For example, SLCT~\cite{slct}, LFA~\cite{lfa}, LogCluster~\cite{logcluster}, and Logram~\cite{logram} try to use different methods to extract frequent patterns in logs.
(2) \textit{Clustering-based} methods aim to group similar logs first, assuming that logs in the same cluster share the same template, and then extract the common tokens as the template in each cluster.
Some clustering-based methods can perform in an online manner because they adopt an online grouping strategy rather than clustering all the offline logs at once.
Specifically, LKE~\cite{lke}, LogSig~\cite{logsig}, LogMine~\cite{logmine} are offline methods, SHISO~\cite{shiso}, and LenMa~\cite{lenma} are online methods.
(3) \textit{Heuristic-based} methods encode expert domain knowledge into general and effective heuristic rules.
For example, AEL~\cite{ael}, IPLoM~\cite{iplom}, Spell~\cite{spell}, and Drain~\cite{drain} utilize different heuristic rules to extract templates from logs.
In particular, Drain~\cite{drain} achieved SOTA in all open-source traditional unsupervised parsers with a parse tree structure to perform log parsing in an online manner.
Based on Drain's architecture, there are multiple updated log parsers:
POP~\cite{pop} improves Drain and provides a parallel implementation on Spark for distributed deployment.
SPINE~\cite{spine} improves Drain and proposes a progressive clustering step for log template correction via human feedback.

\textbf{Supervised log parsers}
Supervised log parsers are a recently emerging class of parsers. 
They typically use deep learning and use data sampled from a target log source to supervise mine patterns in log templates.
For example, UniParser~\cite{uniparser} utilizes a contrast learning strategy to overcome the pattern difference in heterogeneous log sources based on a BiLSTM~\cite{bilstm} model.
SemParser~\cite{semparser} utilizes a special semantic miner for template extraction along with the concept-instance pair using a BiLSTM~\cite{bilstm} model.
Both of the above log parsers require training a neural network from scratch using labeled training data.
Recently, LogPPT~\cite{logppt} has proposed a new paradigm for log parsing, which uses template-free prompt-tuning~\cite{templatefreeprompttuning} to fine-tune the pre-trained language model, RoBERTa~\cite{roberta}.
By mining the semantic information contained in the pre-trained model itself and fine-tuning it with a small amount of log data (\ie using 32 log samples for each dataset), LogPPT is able to achieve the current SOTA performance on multiple datasets and multiple metrics, outperforming all existing log parsers in accuracy and robustness. 
These results show that mining the pre-trained model for log parsing is feasible and effective.

Unlike other log parsers, \methodname doesn't require heuristic rules, handcrafted features, or fine-tuning on specific training data. 
Instead, it generates the target log template by leveraging the ICL ability of LLMs and mining semantics from related log examples in the prompts.

\section{Conclusion}

In conclusion, our paper introduces a new log parsing framework, named \methodname, that utilizes the in-context learning capability of large language models to enhance log parsing effectiveness. 
Our method mines semantics from several log examples provided in a prompt, enabling precise log template generation without requiring model tuning.
Additionally, we design a prompt format to constrain the output and ensure the quality of the generated log templates from LLMs, which we believe can generalize to other ICL applications. 
Through experiments on 16 public datasets using three different metrics, we demonstrate that \methodname outperforms current log parsers by 3.86\%, 20.4\%, and 16.7\%, in terms of PA, PTA, and RTA, respectively.
In a broader sense, we believe this framework has significant potential to improve the effectiveness and practicality of various downstream automated log analysis applications.

\section{Data Availability}
We uploaded our algorithm implementation and datasets in GitHub repository at~\url{https://github.com/logpai/logparser}

\begin{acks}
This paper was supported by the National Natural Science Foundation of China (No. 62102340) and Shenzhen Science and Technology Program.
\end{acks}

\balance
\bibliographystyle{ACM-Reference-Format}
\bibliography{ref}


\begin{thebibliography}{56}


\ifx \showCODEN    \undefined \def \showCODEN     #1{\unskip}     \fi
\ifx \showDOI      \undefined \def \showDOI       #1{#1}\fi
\ifx \showISBNx    \undefined \def \showISBNx     #1{\unskip}     \fi
\ifx \showISBNxiii \undefined \def \showISBNxiii  #1{\unskip}     \fi
\ifx \showISSN     \undefined \def \showISSN      #1{\unskip}     \fi
\ifx \showLCCN     \undefined \def \showLCCN      #1{\unskip}     \fi
\ifx \shownote     \undefined \def \shownote      #1{#1}          \fi
\ifx \showarticletitle \undefined \def \showarticletitle #1{#1}   \fi
\ifx \showURL      \undefined \def \showURL       {\relax}        \fi
\providecommand\bibfield[2]{#2}
\providecommand\bibinfo[2]{#2}
\providecommand\natexlab[1]{#1}
\providecommand\showeprint[2][]{arXiv:#2}

\bibitem[Ahmed et~al\mbox{.}(2023)]%
        {ahmed2023recommending}
\bibfield{author}{\bibinfo{person}{Toufique Ahmed}, \bibinfo{person}{Supriyo
  Ghosh}, \bibinfo{person}{Chetan Bansal}, \bibinfo{person}{Thomas Zimmermann},
  \bibinfo{person}{Xuchao Zhang}, {and} \bibinfo{person}{Saravan Rajmohan}.}
  \bibinfo{year}{2023}\natexlab{}.
\newblock \showarticletitle{Recommending Root-Cause and Mitigation Steps for
  Cloud Incidents using Large Language Models}.
\newblock \bibinfo{journal}{\emph{ICSE}} (\bibinfo{year}{2023}).
\newblock


\bibitem[Amar and Rigby(2019)]%
        {localizelog}
\bibfield{author}{\bibinfo{person}{Anunay Amar} {and} \bibinfo{person}{Peter~C
  Rigby}.} \bibinfo{year}{2019}\natexlab{}.
\newblock \showarticletitle{Mining historical test logs to predict bugs and
  localize faults in the test logs}. In \bibinfo{booktitle}{\emph{2019 IEEE/ACM
  41st International Conference on Software Engineering (ICSE)}}. IEEE,
  \bibinfo{pages}{140--151}.
\newblock


\bibitem[Brown et~al\mbox{.}(2020)]%
        {gpt3}
\bibfield{author}{\bibinfo{person}{Tom Brown}, \bibinfo{person}{Benjamin Mann},
  \bibinfo{person}{Nick Ryder}, \bibinfo{person}{Melanie Subbiah},
  \bibinfo{person}{Jared~D Kaplan}, \bibinfo{person}{Prafulla Dhariwal},
  \bibinfo{person}{Arvind Neelakantan}, \bibinfo{person}{Pranav Shyam},
  \bibinfo{person}{Girish Sastry}, \bibinfo{person}{Amanda Askell},
  {et~al\mbox{.}}} \bibinfo{year}{2020}\natexlab{}.
\newblock \showarticletitle{Language models are few-shot learners}.
\newblock \bibinfo{journal}{\emph{Advances in neural information processing
  systems}}  \bibinfo{volume}{33} (\bibinfo{year}{2020}),
  \bibinfo{pages}{1877--1901}.
\newblock


\bibitem[Chen et~al\mbox{.}(2018)]%
        {dpp}
\bibfield{author}{\bibinfo{person}{Laming Chen}, \bibinfo{person}{Guoxin
  Zhang}, {and} \bibinfo{person}{Eric Zhou}.} \bibinfo{year}{2018}\natexlab{}.
\newblock \showarticletitle{Fast greedy map inference for determinantal point
  process to improve recommendation diversity}.
\newblock \bibinfo{journal}{\emph{Advances in Neural Information Processing
  Systems}}  \bibinfo{volume}{31} (\bibinfo{year}{2018}).
\newblock


\bibitem[Dai et~al\mbox{.}(2020)]%
        {logram}
\bibfield{author}{\bibinfo{person}{Hetong Dai}, \bibinfo{person}{Heng Li},
  \bibinfo{person}{Che~Shao Chen}, \bibinfo{person}{Weiyi Shang}, {and}
  \bibinfo{person}{Tse-Hsun Chen}.} \bibinfo{year}{2020}\natexlab{}.
\newblock \showarticletitle{Logram: Efficient log parsing using n-gram
  dictionaries}.
\newblock \bibinfo{journal}{\emph{IEEE Transactions on Software Engineering}}
  (\bibinfo{year}{2020}).
\newblock


\bibitem[Deng et~al\mbox{.}(2022)]%
        {fuzzer}
\bibfield{author}{\bibinfo{person}{Yinlin Deng},
  \bibinfo{person}{Chunqiu~Steven Xia}, \bibinfo{person}{Haoran Peng},
  \bibinfo{person}{Chenyuan Yang}, {and} \bibinfo{person}{Lingming Zhang}.}
  \bibinfo{year}{2022}\natexlab{}.
\newblock \showarticletitle{Fuzzing Deep-Learning Libraries via Large Language
  Models}.
\newblock \bibinfo{journal}{\emph{arXiv preprint arXiv:2212.14834}}
  (\bibinfo{year}{2022}).
\newblock


\bibitem[Devlin et~al\mbox{.}(2018)]%
        {bert}
\bibfield{author}{\bibinfo{person}{Jacob Devlin}, \bibinfo{person}{Ming-Wei
  Chang}, \bibinfo{person}{Kenton Lee}, {and} \bibinfo{person}{Kristina
  Toutanova}.} \bibinfo{year}{2018}\natexlab{}.
\newblock \showarticletitle{Bert: Pre-training of deep bidirectional
  transformers for language understanding}.
\newblock \bibinfo{journal}{\emph{arXiv preprint arXiv:1810.04805}}
  (\bibinfo{year}{2018}).
\newblock


\bibitem[Ding et~al\mbox{.}(2015)]%
        {log2}
\bibfield{author}{\bibinfo{person}{Rui Ding}, \bibinfo{person}{Hucheng Zhou},
  \bibinfo{person}{Jian-Guang Lou}, \bibinfo{person}{Hongyu Zhang},
  \bibinfo{person}{Qingwei Lin}, \bibinfo{person}{Qiang Fu},
  \bibinfo{person}{Dongmei Zhang}, {and} \bibinfo{person}{Tao Xie}.}
  \bibinfo{year}{2015}\natexlab{}.
\newblock \showarticletitle{Log2: A cost-aware logging mechanism for
  performance diagnosis}. In \bibinfo{booktitle}{\emph{2015 $\{$USENIX$\}$
  Annual Technical Conference ($\{$USENIX$\}$$\{$ATC$\}$ 15)}}.
  \bibinfo{pages}{139--150}.
\newblock


\bibitem[Dong et~al\mbox{.}(2022)]%
        {icl}
\bibfield{author}{\bibinfo{person}{Qingxiu Dong}, \bibinfo{person}{Lei Li},
  \bibinfo{person}{Damai Dai}, \bibinfo{person}{Ce Zheng},
  \bibinfo{person}{Zhiyong Wu}, \bibinfo{person}{Baobao Chang},
  \bibinfo{person}{Xu Sun}, \bibinfo{person}{Jingjing Xu}, {and}
  \bibinfo{person}{Zhifang Sui}.} \bibinfo{year}{2022}\natexlab{}.
\newblock \showarticletitle{A Survey for In-context Learning}.
\newblock \bibinfo{journal}{\emph{arXiv preprint arXiv:2301.00234}}
  (\bibinfo{year}{2022}).
\newblock


\bibitem[Du and Li(2016)]%
        {spell}
\bibfield{author}{\bibinfo{person}{Min Du} {and} \bibinfo{person}{Feifei Li}.}
  \bibinfo{year}{2016}\natexlab{}.
\newblock \showarticletitle{Spell: Streaming parsing of system event logs}. In
  \bibinfo{booktitle}{\emph{2016 IEEE 16th International Conference on Data
  Mining (ICDM)}}. IEEE, \bibinfo{pages}{859--864}.
\newblock


\bibitem[Du et~al\mbox{.}(2017)]%
        {deeplog}
\bibfield{author}{\bibinfo{person}{Min Du}, \bibinfo{person}{Feifei Li},
  \bibinfo{person}{Guineng Zheng}, {and} \bibinfo{person}{Vivek Srikumar}.}
  \bibinfo{year}{2017}\natexlab{}.
\newblock \showarticletitle{Deeplog: Anomaly detection and diagnosis from
  system logs through deep learning}. In \bibinfo{booktitle}{\emph{Proceedings
  of the 2017 ACM SIGSAC conference on computer and communications security}}.
  \bibinfo{pages}{1285--1298}.
\newblock


\bibitem[Fu et~al\mbox{.}(2009a)]%
        {ael}
\bibfield{author}{\bibinfo{person}{Qiang Fu}, \bibinfo{person}{Jian-Guang Lou},
  \bibinfo{person}{Yi Wang}, {and} \bibinfo{person}{Jiang Li}.}
  \bibinfo{year}{2009}\natexlab{a}.
\newblock \showarticletitle{Execution anomaly detection in distributed systems
  through unstructured log analysis}. In \bibinfo{booktitle}{\emph{2009 ninth
  IEEE international conference on data mining}}. IEEE,
  \bibinfo{pages}{149--158}.
\newblock


\bibitem[Fu et~al\mbox{.}(2009b)]%
        {lke}
\bibfield{author}{\bibinfo{person}{Qiang Fu}, \bibinfo{person}{Jian-Guang Lou},
  \bibinfo{person}{Yi Wang}, {and} \bibinfo{person}{Jiang Li}.}
  \bibinfo{year}{2009}\natexlab{b}.
\newblock \showarticletitle{Execution anomaly detection in distributed systems
  through unstructured log analysis}. In \bibinfo{booktitle}{\emph{2009 ninth
  IEEE international conference on data mining}}. IEEE,
  \bibinfo{pages}{149--158}.
\newblock


\bibitem[Graves et~al\mbox{.}(2013)]%
        {bilstm}
\bibfield{author}{\bibinfo{person}{Alex Graves}, \bibinfo{person}{Abdel-rahman
  Mohamed}, {and} \bibinfo{person}{Geoffrey Hinton}.}
  \bibinfo{year}{2013}\natexlab{}.
\newblock \showarticletitle{Speech recognition with deep recurrent neural
  networks}. In \bibinfo{booktitle}{\emph{2013 IEEE international conference on
  acoustics, speech and signal processing}}. Ieee, \bibinfo{pages}{6645--6649}.
\newblock


\bibitem[Hamooni et~al\mbox{.}(2016)]%
        {logmine}
\bibfield{author}{\bibinfo{person}{Hossein Hamooni}, \bibinfo{person}{Biplob
  Debnath}, \bibinfo{person}{Jianwu Xu}, \bibinfo{person}{Hui Zhang},
  \bibinfo{person}{Guofei Jiang}, {and} \bibinfo{person}{Abdullah Mueen}.}
  \bibinfo{year}{2016}\natexlab{}.
\newblock \showarticletitle{Logmine: Fast pattern recognition for log
  analytics}. In \bibinfo{booktitle}{\emph{Proceedings of the 25th ACM
  International on Conference on Information and Knowledge Management}}.
  \bibinfo{pages}{1573--1582}.
\newblock


\bibitem[He et~al\mbox{.}(2016b)]%
        {evaluationstudy}
\bibfield{author}{\bibinfo{person}{Pinjia He}, \bibinfo{person}{Jieming Zhu},
  \bibinfo{person}{Shilin He}, \bibinfo{person}{Jian Li}, {and}
  \bibinfo{person}{Michael~R Lyu}.} \bibinfo{year}{2016}\natexlab{b}.
\newblock \showarticletitle{An evaluation study on log parsing and its use in
  log mining}. In \bibinfo{booktitle}{\emph{2016 46th annual IEEE/IFIP
  international conference on dependable systems and networks (DSN)}}. IEEE,
  \bibinfo{pages}{654--661}.
\newblock


\bibitem[He et~al\mbox{.}(2017a)]%
        {pop}
\bibfield{author}{\bibinfo{person}{Pinjia He}, \bibinfo{person}{Jieming Zhu},
  \bibinfo{person}{Shilin He}, \bibinfo{person}{Jian Li}, {and}
  \bibinfo{person}{Michael~R Lyu}.} \bibinfo{year}{2017}\natexlab{a}.
\newblock \showarticletitle{Towards automated log parsing for large-scale log
  data analysis}.
\newblock \bibinfo{journal}{\emph{IEEE Transactions on Dependable and Secure
  Computing}} \bibinfo{volume}{15}, \bibinfo{number}{6} (\bibinfo{year}{2017}),
  \bibinfo{pages}{931--944}.
\newblock


\bibitem[He et~al\mbox{.}(2017b)]%
        {drain}
\bibfield{author}{\bibinfo{person}{Pinjia He}, \bibinfo{person}{Jieming Zhu},
  \bibinfo{person}{Zibin Zheng}, {and} \bibinfo{person}{Michael~R Lyu}.}
  \bibinfo{year}{2017}\natexlab{b}.
\newblock \showarticletitle{Drain: An online log parsing approach with fixed
  depth tree}. In \bibinfo{booktitle}{\emph{2017 IEEE international conference
  on web services (ICWS)}}. IEEE, \bibinfo{pages}{33--40}.
\newblock


\bibitem[He et~al\mbox{.}(2021)]%
        {survey}
\bibfield{author}{\bibinfo{person}{Shilin He}, \bibinfo{person}{Pinjia He},
  \bibinfo{person}{Zhuangbin Chen}, \bibinfo{person}{Tianyi Yang},
  \bibinfo{person}{Yuxin Su}, {and} \bibinfo{person}{Michael~R Lyu}.}
  \bibinfo{year}{2021}\natexlab{}.
\newblock \showarticletitle{A survey on automated log analysis for reliability
  engineering}.
\newblock \bibinfo{journal}{\emph{ACM computing surveys (CSUR)}}
  \bibinfo{volume}{54}, \bibinfo{number}{6} (\bibinfo{year}{2021}),
  \bibinfo{pages}{1--37}.
\newblock


\bibitem[He et~al\mbox{.}(2018)]%
        {idlogm}
\bibfield{author}{\bibinfo{person}{Shilin He}, \bibinfo{person}{Qingwei Lin},
  \bibinfo{person}{Jian-Guang Lou}, \bibinfo{person}{Hongyu Zhang},
  \bibinfo{person}{Michael~R Lyu}, {and} \bibinfo{person}{Dongmei Zhang}.}
  \bibinfo{year}{2018}\natexlab{}.
\newblock \showarticletitle{Identifying impactful service system problems via
  log analysis}. In \bibinfo{booktitle}{\emph{Proceedings of the 2018 26th ACM
  Joint Meeting on European Software Engineering Conference and Symposium on
  the Foundations of Software Engineering}}. \bibinfo{pages}{60--70}.
\newblock


\bibitem[He et~al\mbox{.}(2016a)]%
        {exp}
\bibfield{author}{\bibinfo{person}{Shilin He}, \bibinfo{person}{Jieming Zhu},
  \bibinfo{person}{Pinjia He}, {and} \bibinfo{person}{Michael~R Lyu}.}
  \bibinfo{year}{2016}\natexlab{a}.
\newblock \showarticletitle{Experience report: System log analysis for anomaly
  detection}. In \bibinfo{booktitle}{\emph{2016 IEEE 27th international
  symposium on software reliability engineering (ISSRE)}}. IEEE,
  \bibinfo{pages}{207--218}.
\newblock


\bibitem[Huo et~al\mbox{.}(2021)]%
        {semparser}
\bibfield{author}{\bibinfo{person}{Yintong Huo}, \bibinfo{person}{Yuxin Su},
  \bibinfo{person}{Baitong Li}, {and} \bibinfo{person}{Michael~R Lyu}.}
  \bibinfo{year}{2021}\natexlab{}.
\newblock \showarticletitle{SemParser: A Semantic Parser for Log Analysis}.
\newblock \bibinfo{journal}{\emph{arXiv preprint arXiv:2112.12636}}
  (\bibinfo{year}{2021}).
\newblock


\bibitem[Khan et~al\mbox{.}(2022)]%
        {guideline}
\bibfield{author}{\bibinfo{person}{Zanis~Ali Khan}, \bibinfo{person}{Donghwan
  Shin}, \bibinfo{person}{Domenico Bianculli}, {and} \bibinfo{person}{Lionel
  Briand}.} \bibinfo{year}{2022}\natexlab{}.
\newblock \showarticletitle{Guidelines for Assessing the Accuracy of Log
  Message Template Identification Techniques}. In
  \bibinfo{booktitle}{\emph{Proceedings of the 44th International Conference on
  Software Engineering (ICSE’22)}}. ACM.
\newblock


\bibitem[Kumar et~al\mbox{.}(2022)]%
        {finetuneharmgeneral}
\bibfield{author}{\bibinfo{person}{Ananya Kumar}, \bibinfo{person}{Aditi
  Raghunathan}, \bibinfo{person}{Robbie Jones}, \bibinfo{person}{Tengyu Ma},
  {and} \bibinfo{person}{Percy Liang}.} \bibinfo{year}{2022}\natexlab{}.
\newblock \showarticletitle{Fine-tuning can distort pretrained features and
  underperform out-of-distribution}.
\newblock \bibinfo{journal}{\emph{arXiv preprint arXiv:2202.10054}}
  (\bibinfo{year}{2022}).
\newblock


\bibitem[Le and Zhang(2023a)]%
        {logppt}
\bibfield{author}{\bibinfo{person}{Van-Hoang Le} {and} \bibinfo{person}{Hongyu
  Zhang}.} \bibinfo{year}{2023}\natexlab{a}.
\newblock \showarticletitle{Log Parsing with Prompt-based Few-shot Learning}.
\newblock \bibinfo{journal}{\emph{arXiv preprint arXiv:2302.07435}}
  (\bibinfo{year}{2023}).
\newblock


\bibitem[Le and Zhang(2023b)]%
        {logpptrepo}
\bibfield{author}{\bibinfo{person}{Van-Hoang Le} {and} \bibinfo{person}{Hongyu
  Zhang}.} \bibinfo{year}{2023}\natexlab{b}.
\newblock \bibinfo{title}{{Repository of LogPPT}}.
\newblock \bibinfo{howpublished}{\url{https://github.com/LogIntelligence/LogPPT
  }}.
\newblock
\newblock
\shownote{[Online; accessed 19-March-2023]}.


\bibitem[Lewis et~al\mbox{.}(2019)]%
        {bart}
\bibfield{author}{\bibinfo{person}{Mike Lewis}, \bibinfo{person}{Yinhan Liu},
  \bibinfo{person}{Naman Goyal}, \bibinfo{person}{Marjan Ghazvininejad},
  \bibinfo{person}{Abdelrahman Mohamed}, \bibinfo{person}{Omer Levy},
  \bibinfo{person}{Ves Stoyanov}, {and} \bibinfo{person}{Luke Zettlemoyer}.}
  \bibinfo{year}{2019}\natexlab{}.
\newblock \showarticletitle{Bart: Denoising sequence-to-sequence pre-training
  for natural language generation, translation, and comprehension}.
\newblock \bibinfo{journal}{\emph{arXiv preprint arXiv:1910.13461}}
  (\bibinfo{year}{2019}).
\newblock


\bibitem[Li et~al\mbox{.}(2022)]%
        {auger}
\bibfield{author}{\bibinfo{person}{Lingwei Li}, \bibinfo{person}{Li Yang},
  \bibinfo{person}{Huaxi Jiang}, \bibinfo{person}{Jun Yan},
  \bibinfo{person}{Tiejian Luo}, \bibinfo{person}{Zihan Hua},
  \bibinfo{person}{Geng Liang}, {and} \bibinfo{person}{Chun Zuo}.}
  \bibinfo{year}{2022}\natexlab{}.
\newblock \showarticletitle{AUGER: automatically generating review comments
  with pre-training models}. In \bibinfo{booktitle}{\emph{Proceedings of the
  30th ACM Joint European Software Engineering Conference and Symposium on the
  Foundations of Software Engineering}}. \bibinfo{pages}{1009--1021}.
\newblock


\bibitem[Lin et~al\mbox{.}(2016)]%
        {idlogc}
\bibfield{author}{\bibinfo{person}{Qingwei Lin}, \bibinfo{person}{Hongyu
  Zhang}, \bibinfo{person}{Jian-Guang Lou}, \bibinfo{person}{Yu Zhang}, {and}
  \bibinfo{person}{Xuewei Chen}.} \bibinfo{year}{2016}\natexlab{}.
\newblock \showarticletitle{Log clustering based problem identification for
  online service systems}. In \bibinfo{booktitle}{\emph{2016 IEEE/ACM 38th
  International Conference on Software Engineering Companion (ICSE-C)}}. IEEE,
  \bibinfo{pages}{102--111}.
\newblock


\bibitem[Liu et~al\mbox{.}(2021)]%
        {kate}
\bibfield{author}{\bibinfo{person}{Jiachang Liu}, \bibinfo{person}{Dinghan
  Shen}, \bibinfo{person}{Yizhe Zhang}, \bibinfo{person}{Bill Dolan},
  \bibinfo{person}{Lawrence Carin}, {and} \bibinfo{person}{Weizhu Chen}.}
  \bibinfo{year}{2021}\natexlab{}.
\newblock \showarticletitle{What Makes Good In-Context Examples for GPT-$3 $?}
\newblock \bibinfo{journal}{\emph{arXiv preprint arXiv:2101.06804}}
  (\bibinfo{year}{2021}).
\newblock


\bibitem[Liu et~al\mbox{.}(2023)]%
        {promptsurvey}
\bibfield{author}{\bibinfo{person}{Pengfei Liu}, \bibinfo{person}{Weizhe Yuan},
  \bibinfo{person}{Jinlan Fu}, \bibinfo{person}{Zhengbao Jiang},
  \bibinfo{person}{Hiroaki Hayashi}, {and} \bibinfo{person}{Graham Neubig}.}
  \bibinfo{year}{2023}\natexlab{}.
\newblock \showarticletitle{Pre-train, prompt, and predict: A systematic survey
  of prompting methods in natural language processing}.
\newblock \bibinfo{journal}{\emph{Comput. Surveys}} \bibinfo{volume}{55},
  \bibinfo{number}{9} (\bibinfo{year}{2023}), \bibinfo{pages}{1--35}.
\newblock


\bibitem[Liu et~al\mbox{.}(2019)]%
        {roberta}
\bibfield{author}{\bibinfo{person}{Yinhan Liu}, \bibinfo{person}{Myle Ott},
  \bibinfo{person}{Naman Goyal}, \bibinfo{person}{Jingfei Du},
  \bibinfo{person}{Mandar Joshi}, \bibinfo{person}{Danqi Chen},
  \bibinfo{person}{Omer Levy}, \bibinfo{person}{Mike Lewis},
  \bibinfo{person}{Luke Zettlemoyer}, {and} \bibinfo{person}{Veselin
  Stoyanov}.} \bibinfo{year}{2019}\natexlab{}.
\newblock \showarticletitle{Roberta: A robustly optimized bert pretraining
  approach}.
\newblock \bibinfo{journal}{\emph{arXiv preprint arXiv:1907.11692}}
  (\bibinfo{year}{2019}).
\newblock


\bibitem[Liu et~al\mbox{.}(2022)]%
        {uniparser}
\bibfield{author}{\bibinfo{person}{Yudong Liu}, \bibinfo{person}{Xu Zhang},
  \bibinfo{person}{Shilin He}, \bibinfo{person}{Hongyu Zhang},
  \bibinfo{person}{Liqun Li}, \bibinfo{person}{Yu Kang}, \bibinfo{person}{Yong
  Xu}, \bibinfo{person}{Minghua Ma}, \bibinfo{person}{Qingwei Lin},
  \bibinfo{person}{Yingnong Dang}, {et~al\mbox{.}}}
  \bibinfo{year}{2022}\natexlab{}.
\newblock \showarticletitle{UniParser: A Unified Log Parser for Heterogeneous
  Log Data}. In \bibinfo{booktitle}{\emph{Proceedings of the ACM Web Conference
  2022}}. \bibinfo{pages}{1893--1901}.
\newblock


\bibitem[Lou et~al\mbox{.}(2010)]%
        {mininglog}
\bibfield{author}{\bibinfo{person}{Jian-Guang Lou}, \bibinfo{person}{Qiang Fu},
  \bibinfo{person}{Shenqi Yang}, \bibinfo{person}{Ye Xu}, {and}
  \bibinfo{person}{Jiang Li}.} \bibinfo{year}{2010}\natexlab{}.
\newblock \showarticletitle{Mining invariants from console logs for system
  problem detection}. In \bibinfo{booktitle}{\emph{2010 USENIX Annual Technical
  Conference (USENIX ATC 10)}}.
\newblock


\bibitem[Ma et~al\mbox{.}(2021)]%
        {templatefreeprompttuning}
\bibfield{author}{\bibinfo{person}{Ruotian Ma}, \bibinfo{person}{Xin Zhou},
  \bibinfo{person}{Tao Gui}, \bibinfo{person}{Yiding Tan},
  \bibinfo{person}{Linyang Li}, \bibinfo{person}{Qi Zhang}, {and}
  \bibinfo{person}{Xuanjing Huang}.} \bibinfo{year}{2021}\natexlab{}.
\newblock \showarticletitle{Template-free prompt tuning for few-shot NER}.
\newblock \bibinfo{journal}{\emph{arXiv preprint arXiv:2109.13532}}
  (\bibinfo{year}{2021}).
\newblock


\bibitem[Makanju et~al\mbox{.}(2009)]%
        {iplom}
\bibfield{author}{\bibinfo{person}{Adetokunbo~AO Makanju},
  \bibinfo{person}{A~Nur Zincir-Heywood}, {and} \bibinfo{person}{Evangelos~E
  Milios}.} \bibinfo{year}{2009}\natexlab{}.
\newblock \showarticletitle{Clustering event logs using iterative
  partitioning}. In \bibinfo{booktitle}{\emph{Proceedings of the 15th ACM
  SIGKDD international conference on Knowledge discovery and data mining}}.
  \bibinfo{pages}{1255--1264}.
\newblock


\bibitem[Mizutani(2013)]%
        {shiso}
\bibfield{author}{\bibinfo{person}{Masayoshi Mizutani}.}
  \bibinfo{year}{2013}\natexlab{}.
\newblock \showarticletitle{Incremental mining of system log format}. In
  \bibinfo{booktitle}{\emph{2013 IEEE International Conference on Services
  Computing}}. IEEE, \bibinfo{pages}{595--602}.
\newblock


\bibitem[Nagappan and Vouk(2010)]%
        {lfa}
\bibfield{author}{\bibinfo{person}{Meiyappan Nagappan} {and}
  \bibinfo{person}{Mladen~A Vouk}.} \bibinfo{year}{2010}\natexlab{}.
\newblock \showarticletitle{Abstracting log lines to log event types for mining
  software system logs}. In \bibinfo{booktitle}{\emph{2010 7th IEEE Working
  Conference on Mining Software Repositories (MSR 2010)}}. IEEE,
  \bibinfo{pages}{114--117}.
\newblock


\bibitem[Radford et~al\mbox{.}(2018)]%
        {gpt}
\bibfield{author}{\bibinfo{person}{Alec Radford}, \bibinfo{person}{Karthik
  Narasimhan}, \bibinfo{person}{Tim Salimans}, \bibinfo{person}{Ilya
  Sutskever}, {et~al\mbox{.}}} \bibinfo{year}{2018}\natexlab{}.
\newblock \showarticletitle{Improving language understanding by generative
  pre-training}.
\newblock  (\bibinfo{year}{2018}).
\newblock


\bibitem[Radford et~al\mbox{.}(2019)]%
        {gpt2}
\bibfield{author}{\bibinfo{person}{Alec Radford}, \bibinfo{person}{Jeffrey Wu},
  \bibinfo{person}{Rewon Child}, \bibinfo{person}{David Luan},
  \bibinfo{person}{Dario Amodei}, \bibinfo{person}{Ilya Sutskever},
  {et~al\mbox{.}}} \bibinfo{year}{2019}\natexlab{}.
\newblock \showarticletitle{Language models are unsupervised multitask
  learners}.
\newblock \bibinfo{journal}{\emph{OpenAI blog}} \bibinfo{volume}{1},
  \bibinfo{number}{8} (\bibinfo{year}{2019}), \bibinfo{pages}{9}.
\newblock


\bibitem[Raffel et~al\mbox{.}(2020)]%
        {t5}
\bibfield{author}{\bibinfo{person}{Colin Raffel}, \bibinfo{person}{Noam
  Shazeer}, \bibinfo{person}{Adam Roberts}, \bibinfo{person}{Katherine Lee},
  \bibinfo{person}{Sharan Narang}, \bibinfo{person}{Michael Matena},
  \bibinfo{person}{Yanqi Zhou}, \bibinfo{person}{Wei Li}, {and}
  \bibinfo{person}{Peter~J. Liu}.} \bibinfo{year}{2020}\natexlab{}.
\newblock \showarticletitle{Exploring the Limits of Transfer Learning with a
  Unified Text-to-Text Transformer}.
\newblock \bibinfo{journal}{\emph{J. Mach. Learn. Res.}} \bibinfo{volume}{21},
  \bibinfo{number}{1}, Article \bibinfo{articleno}{140} (\bibinfo{date}{jan}
  \bibinfo{year}{2020}), \bibinfo{numpages}{67}~pages.
\newblock
\showISSN{1532-4435}


\bibitem[Rubin et~al\mbox{.}(2021)]%
        {epr}
\bibfield{author}{\bibinfo{person}{Ohad Rubin}, \bibinfo{person}{Jonathan
  Herzig}, {and} \bibinfo{person}{Jonathan Berant}.}
  \bibinfo{year}{2021}\natexlab{}.
\newblock \showarticletitle{Learning to retrieve prompts for in-context
  learning}.
\newblock \bibinfo{journal}{\emph{arXiv preprint arXiv:2112.08633}}
  (\bibinfo{year}{2021}).
\newblock


\bibitem[Shang et~al\mbox{.}(2013)]%
        {logverify}
\bibfield{author}{\bibinfo{person}{Weiyi Shang}, \bibinfo{person}{Zhen~Ming
  Jiang}, \bibinfo{person}{Hadi Hemmati}, \bibinfo{person}{Brain Adams},
  \bibinfo{person}{Ahmed~E Hassan}, {and} \bibinfo{person}{Patrick Martin}.}
  \bibinfo{year}{2013}\natexlab{}.
\newblock \showarticletitle{Assisting developers of big data analytics
  applications when deploying on hadoop clouds}. In
  \bibinfo{booktitle}{\emph{2013 35th International Conference on Software
  Engineering (ICSE)}}. IEEE, \bibinfo{pages}{402--411}.
\newblock


\bibitem[Shima(2016)]%
        {lenma}
\bibfield{author}{\bibinfo{person}{Keiichi Shima}.}
  \bibinfo{year}{2016}\natexlab{}.
\newblock \showarticletitle{Length matters: Clustering system log messages
  using length of words}.
\newblock \bibinfo{journal}{\emph{arXiv preprint arXiv:1611.03213}}
  (\bibinfo{year}{2016}).
\newblock


\bibitem[Tang et~al\mbox{.}(2011)]%
        {logsig}
\bibfield{author}{\bibinfo{person}{Liang Tang}, \bibinfo{person}{Tao Li}, {and}
  \bibinfo{person}{Chang-Shing Perng}.} \bibinfo{year}{2011}\natexlab{}.
\newblock \showarticletitle{LogSig: Generating system events from raw textual
  logs}. In \bibinfo{booktitle}{\emph{Proceedings of the 20th ACM international
  conference on Information and knowledge management}}.
  \bibinfo{pages}{785--794}.
\newblock


\bibitem[Vaarandi(2003)]%
        {slct}
\bibfield{author}{\bibinfo{person}{Risto Vaarandi}.}
  \bibinfo{year}{2003}\natexlab{}.
\newblock \showarticletitle{A data clustering algorithm for mining patterns
  from event logs}. In \bibinfo{booktitle}{\emph{Proceedings of the 3rd IEEE
  Workshop on IP Operations \& Management (IPOM 2003)(IEEE Cat. No. 03EX764)}}.
  Ieee, \bibinfo{pages}{119--126}.
\newblock


\bibitem[Vaarandi and Pihelgas(2015)]%
        {logcluster}
\bibfield{author}{\bibinfo{person}{Risto Vaarandi} {and} \bibinfo{person}{Mauno
  Pihelgas}.} \bibinfo{year}{2015}\natexlab{}.
\newblock \showarticletitle{Logcluster-a data clustering and pattern mining
  algorithm for event logs}. In \bibinfo{booktitle}{\emph{2015 11th
  International conference on network and service management (CNSM)}}. IEEE,
  \bibinfo{pages}{1--7}.
\newblock


\bibitem[Wang et~al\mbox{.}(2022a)]%
        {nomoreft}
\bibfield{author}{\bibinfo{person}{Chaozheng Wang}, \bibinfo{person}{Yuanhang
  Yang}, \bibinfo{person}{Cuiyun Gao}, \bibinfo{person}{Yun Peng},
  \bibinfo{person}{Hongyu Zhang}, {and} \bibinfo{person}{Michael~R Lyu}.}
  \bibinfo{year}{2022}\natexlab{a}.
\newblock \showarticletitle{No more fine-tuning? an experimental evaluation of
  prompt tuning in code intelligence}. In \bibinfo{booktitle}{\emph{Proceedings
  of the 30th ACM Joint European Software Engineering Conference and Symposium
  on the Foundations of Software Engineering}}. \bibinfo{pages}{382--394}.
\newblock


\bibitem[Wang et~al\mbox{.}(2022b)]%
        {spine}
\bibfield{author}{\bibinfo{person}{Xuheng Wang}, \bibinfo{person}{Xu Zhang},
  \bibinfo{person}{Liqun Li}, \bibinfo{person}{Shilin He},
  \bibinfo{person}{Hongyu Zhang}, \bibinfo{person}{Yudong Liu},
  \bibinfo{person}{Lingling Zheng}, \bibinfo{person}{Yu Kang},
  \bibinfo{person}{Qingwei Lin}, \bibinfo{person}{Yingnong Dang},
  {et~al\mbox{.}}} \bibinfo{year}{2022}\natexlab{b}.
\newblock \showarticletitle{SPINE: a scalable log parser with feedback
  guidance}. In \bibinfo{booktitle}{\emph{Proceedings of the 30th ACM Joint
  European Software Engineering Conference and Symposium on the Foundations of
  Software Engineering}}. \bibinfo{pages}{1198--1208}.
\newblock


\bibitem[Wei et~al\mbox{.}(2022)]%
        {chainofthought}
\bibfield{author}{\bibinfo{person}{Jason Wei}, \bibinfo{person}{Xuezhi Wang},
  \bibinfo{person}{Dale Schuurmans}, \bibinfo{person}{Maarten Bosma},
  \bibinfo{person}{Ed Chi}, \bibinfo{person}{Quoc Le}, {and}
  \bibinfo{person}{Denny Zhou}.} \bibinfo{year}{2022}\natexlab{}.
\newblock \showarticletitle{Chain of thought prompting elicits reasoning in
  large language models}.
\newblock \bibinfo{journal}{\emph{arXiv preprint arXiv:2201.11903}}
  (\bibinfo{year}{2022}).
\newblock


\bibitem[Xia et~al\mbox{.}(2023)]%
        {aprllm}
\bibfield{author}{\bibinfo{person}{Chunqiu~Steven Xia}, \bibinfo{person}{Yifeng
  Ding}, {and} \bibinfo{person}{Lingming Zhang}.}
  \bibinfo{year}{2023}\natexlab{}.
\newblock \showarticletitle{Revisiting the Plastic Surgery Hypothesis via Large
  Language Models}.
\newblock \bibinfo{journal}{\emph{arXiv preprint arXiv:2303.10494}}
  (\bibinfo{year}{2023}).
\newblock


\bibitem[Xia and Zhang(2023)]%
        {aprchatgpt}
\bibfield{author}{\bibinfo{person}{Chunqiu~Steven Xia} {and}
  \bibinfo{person}{Lingming Zhang}.} \bibinfo{year}{2023}\natexlab{}.
\newblock \showarticletitle{Conversational automated program repair}.
\newblock \bibinfo{journal}{\emph{arXiv preprint arXiv:2301.13246}}
  (\bibinfo{year}{2023}).
\newblock


\bibitem[Zhang et~al\mbox{.}(2019)]%
        {robustlog}
\bibfield{author}{\bibinfo{person}{Xu Zhang}, \bibinfo{person}{Yong Xu},
  \bibinfo{person}{Qingwei Lin}, \bibinfo{person}{Bo Qiao},
  \bibinfo{person}{Hongyu Zhang}, \bibinfo{person}{Yingnong Dang},
  \bibinfo{person}{Chunyu Xie}, \bibinfo{person}{Xinsheng Yang},
  \bibinfo{person}{Qian Cheng}, \bibinfo{person}{Ze Li}, {et~al\mbox{.}}}
  \bibinfo{year}{2019}\natexlab{}.
\newblock \showarticletitle{Robust log-based anomaly detection on unstable log
  data}. In \bibinfo{booktitle}{\emph{Proceedings of the 2019 27th ACM Joint
  Meeting on European Software Engineering Conference and Symposium on the
  Foundations of Software Engineering}}. \bibinfo{pages}{807--817}.
\newblock


\bibitem[Zhao et~al\mbox{.}(2021)]%
        {calibrate}
\bibfield{author}{\bibinfo{person}{Zihao Zhao}, \bibinfo{person}{Eric Wallace},
  \bibinfo{person}{Shi Feng}, \bibinfo{person}{Dan Klein}, {and}
  \bibinfo{person}{Sameer Singh}.} \bibinfo{year}{2021}\natexlab{}.
\newblock \showarticletitle{Calibrate before use: Improving few-shot
  performance of language models}. In \bibinfo{booktitle}{\emph{International
  Conference on Machine Learning}}. PMLR, \bibinfo{pages}{12697--12706}.
\newblock


\bibitem[Zhu et~al\mbox{.}({[n.\,d.]})]%
        {benchmarkrepo}
\bibfield{author}{\bibinfo{person}{Jieming Zhu}, \bibinfo{person}{Shilin He},
  \bibinfo{person}{Jinyang Liu}, \bibinfo{person}{Pinjia He},
  \bibinfo{person}{Qi Xie}, \bibinfo{person}{Zibin Zheng}, {and}
  \bibinfo{person}{Michael~R Lyu}.} \bibinfo{year}{[n.\,d.]}\natexlab{}.
\newblock \bibinfo{title}{{Repository of LogPAI}}.
\newblock \bibinfo{howpublished}{\url{https://github.com/logpai/loghub }}.
\newblock


\bibitem[Zhu et~al\mbox{.}(2019)]%
        {benchmark}
\bibfield{author}{\bibinfo{person}{Jieming Zhu}, \bibinfo{person}{Shilin He},
  \bibinfo{person}{Jinyang Liu}, \bibinfo{person}{Pinjia He},
  \bibinfo{person}{Qi Xie}, \bibinfo{person}{Zibin Zheng}, {and}
  \bibinfo{person}{Michael~R Lyu}.} \bibinfo{year}{2019}\natexlab{}.
\newblock \showarticletitle{Tools and benchmarks for automated log parsing}. In
  \bibinfo{booktitle}{\emph{2019 IEEE/ACM 41st International Conference on
  Software Engineering: Software Engineering in Practice (ICSE-SEIP)}}. IEEE,
  \bibinfo{pages}{121--130}.
\newblock


\end{thebibliography}

\end{document}